\documentclass[%
aps,
prx,
reprint,
superscriptaddress,
nofootinbib,
amsmath,
amssymb,
dvipsnames
]{revtex4-2}

\usepackage{
    physics,
    graphicx,
    multirow,
    tabularx,
    mathtools, 
    placeins, 
    nicefrac, 
    hyperref, 
    threeparttable, 
    enumitem,
    makecell,
    adjustbox,
    comment,
    ragged2e,
}
\usepackage{array}

\usepackage[compat=0.3]{yquant}
\usepackage{quantikz}
\useyquantlanguage{groups}

\usepackage[capitalize]{cleveref}
\Crefname{section}{Sec.}{Secs.}

\hypersetup{
    colorlinks,
    linkcolor={blue!50!black},
    citecolor={blue!50!black},
    urlcolor={blue!80!black}
}

\usepackage[caption=false]{subfig} 
\newcommand\black[1]{{\color{black}#1}}

\begin{document}
\title{A Folded Surface Code Architecture for 2D Quantum Hardware}

\newcommand{\qmaddress}{\affiliation{Quantum Motion, 9 Sterling Way, London N7 9HJ, United Kingdom}}
\newcommand{\oxddress}{\affiliation{Department of Materials, University of Oxford, Parks Road, Oxford OX1 3PH, United Kingdom}}
\newcommand{\mathinst}{\affiliation{Mathematical Institute, University of Oxford, Woodstock Road, Oxford OX2 6GG, United Kingdom}}

\newcommand{\oxengaddress}{\affiliation{Department of Engineering Science, University of Oxford, Parks Road, Oxford OX1 3PJ, United Kingdom}}

\author{Zhu Sun}
\email{zhu.sun@exeter.ox.ac.uk}
\oxddress
\mathinst
\qmaddress

\author{Zhenyu Cai}
\email{zhenyu.cai@eng.ox.ac.uk}
\oxengaddress
\qmaddress

\begin{abstract}
    
    Qubit shuttling has become an indispensable ingredient for scaling leading quantum computing platforms, including semiconductor spin, neutral-atom, and trapped-ion qubits, enabling both crosstalk reduction and tighter integration of control hardware. Ref.~\cite{PRXQuantum.4.020345} proposed a scalable architecture that employs short-range shuttling to realize effective three-dimensional connectivity on a strictly two-dimensional device. Building on recent advances in quantum error correction, we show that this architecture enables the native implementation of folded surface codes on 2D hardware, reducing the runtime of all single-qubit logical Clifford gates and logical CNOTs within subsets of qubits from $\order{d}$ in conventional surface code lattice surgery to constant time. We present explicit protocols for these operations and demonstrate that access to a transversal $S$ gate reduces the spacetime volume of 8T-to-CCZ magic-state distillation by more than an order of magnitude compared with standard 2D lattice surgery approaches. Finally, we introduce a new ``virtual-stack" layout that more efficiently exploits the quasi-three-dimensional structure of the architecture, enabling efficient multilayer routing on these two-dimensional devices.
\end{abstract}

\maketitle

\section{Introduction}

In the past three decades, numerous attempts have been made to realize the abstract concept of a quantum computer using a variety of physical modalities. While some leading platforms like superconducting qubits~\cite{nakamura1999coherent,you2011atomic,arute2019quantum,google2025quantum} are composed of a grid of qubits with fixed position in space, qubit shuttling serves as the cornerstone in many other platforms such as trapped-ion \cite{RevModPhys.75.281,blatt2008entangled,kaushal2020shuttling,PhysRevX.12.011032}, neutral atom arrays \cite{bluvstein2022quantum,evered2023high,bluvstein2024logical} and semiconductor spin-qubit \cite{loss1998quantum,buonacorsi2019network,PhysRevApplied.18.024053,PhysRevApplied.18.044064} systems. It is often necessary to increase qubit spacing in these platforms during device design to reduce crosstalk and sometimes to accommodate classical control hardware. Consequently, shuttling becomes a practical necessity for bringing qubits into proximity to perform e.g., entangling operations. However, the capability of shuttling does not automatically correspond to all-to-all connectivity since naive shuttling schemes are not scalable to large array of qubits. While conventional shuttling tracks are generally linear, it was proposed in ref.~\cite{PRXQuantum.4.020345} that, by utilizing shuttling loops in combination with careful pipelining, one can realize a stack of two-dimensional qubit arrays, thereby enabling \emph{scalable} 3D connectivity on a strictly 2D platform. When both inter-loop and intra-loop qubit interactions are available, the system can not only execute standard quantum error correction (QEC) circuits such as stabilizer checks and lattice surgery, but also support transversal two-qubit gates like CNOT. However, the $H$ and $S$ gate would still require standard lattice surgery to implement, which becomes the bottleneck of the Clifford operations. In addition, the detailed implementation of transversal two-qubit gates in the architecture in ref.~\cite{PRXQuantum.4.020345} has not yet been studied.

In parallel, recent advances in QEC have led to a deeper understanding of the stabilizer check circuits for the rotated surface code. It was shown in ref.~\cite{McEwen2023relaxinghardware} that the rotated surface code patch temporarily becomes an unrotated surface code patch during certain stages of the check circuits. Building on this observation, ref.~\cite{chen2024transversal} combines it with a transversal gate scheme for the unrotated surface code to realize, for the first time, fully transversal implementations of all three logical Clifford gates, $\{H, S, CNOT\}$, on the rotated surface code. This method relies on long-range connectivity, which is provided, for example, by neutral-atom platforms. However, as mentioned above, the challenge is to implement such connectivity in a scalable way. In this paper, we show how this can be achieved using a looped pipeline architecture with short-range qubit shuttling by leveraging folded surface code.

To demonstrate the efficiency of our scheme and also for achieving universal computation, we study the implementation of 8T-to-CCZ distillation factory \cite{jones2013low} augmented with magic state cultivation \cite{gidney2024magic}, using a variant of the circuit in ref.~\cite{fazio2025low}. Using looped pipeline architecture with the rotated surface code already achieves a substantial reduction in overhead compared to conventional 8T-to-CCZ factories~\cite{fazio2025low}. By further adopting the folded surface code within our architecture, the availability of a transversal $S$ gate enables an additional reduction of approximately a factor of 2.6 in spacetime volume.

Regarding the layout of logical code patches, we propose a new scheme that leverages the virtual stack structure intrinsic to our architecture. Conceptually, the entire computing region can be viewed as a three-dimensional multilayer structure implemented on a strictly two-dimensional hardware platform, with each layer’s routing space allocated according to the computational tasks performed within it (see \cref{fig:layers} for an overview). We classify large-scale computational tasks into four categories: storage, short-range operations, mid-range operations, and long-range operations. Transitions between layers are enabled by low-cost transversal SWAP gates. We view flexibility as the primary advantage of this layout scheme, and note that the layout proposed in ref.~\cite{PRXQuantum.4.020345} arises as a special case within this more general framework.

The remainder of this manuscript is organized as follows. In \cref{Sec:Background}, we review relevant background on surface codes and their variants, as well as the looped pipeline architecture. Next, in \cref{Sec:rearrange}, we present our methods for implementing intra-loop interactions. In \cref{Sec:folded}, we describe the protocol for performing transversal Clifford gates on folded surface codes within our architecture, and in \cref{Sec:magic} we use these results to construct a magic-state factory for the CCZ state. We then introduce a new layout scheme for logical code patches in \cref{Sec:architecture}, and conclude in \cref{Sec:Conclusion}. \cref{appx:runtimes} presents the runtime analysis, while \cref{appx:inter-loop} outlines a potential architectural improvement requiring further hardware advances.

\section{Background}
\label{Sec:Background}
\subsection{Surface codes and their variants}

\begin{figure}[b]
    \centering
    \includegraphics[width=0.9\linewidth, trim=0cm 0.6cm 0cm 0.6cm, clip]{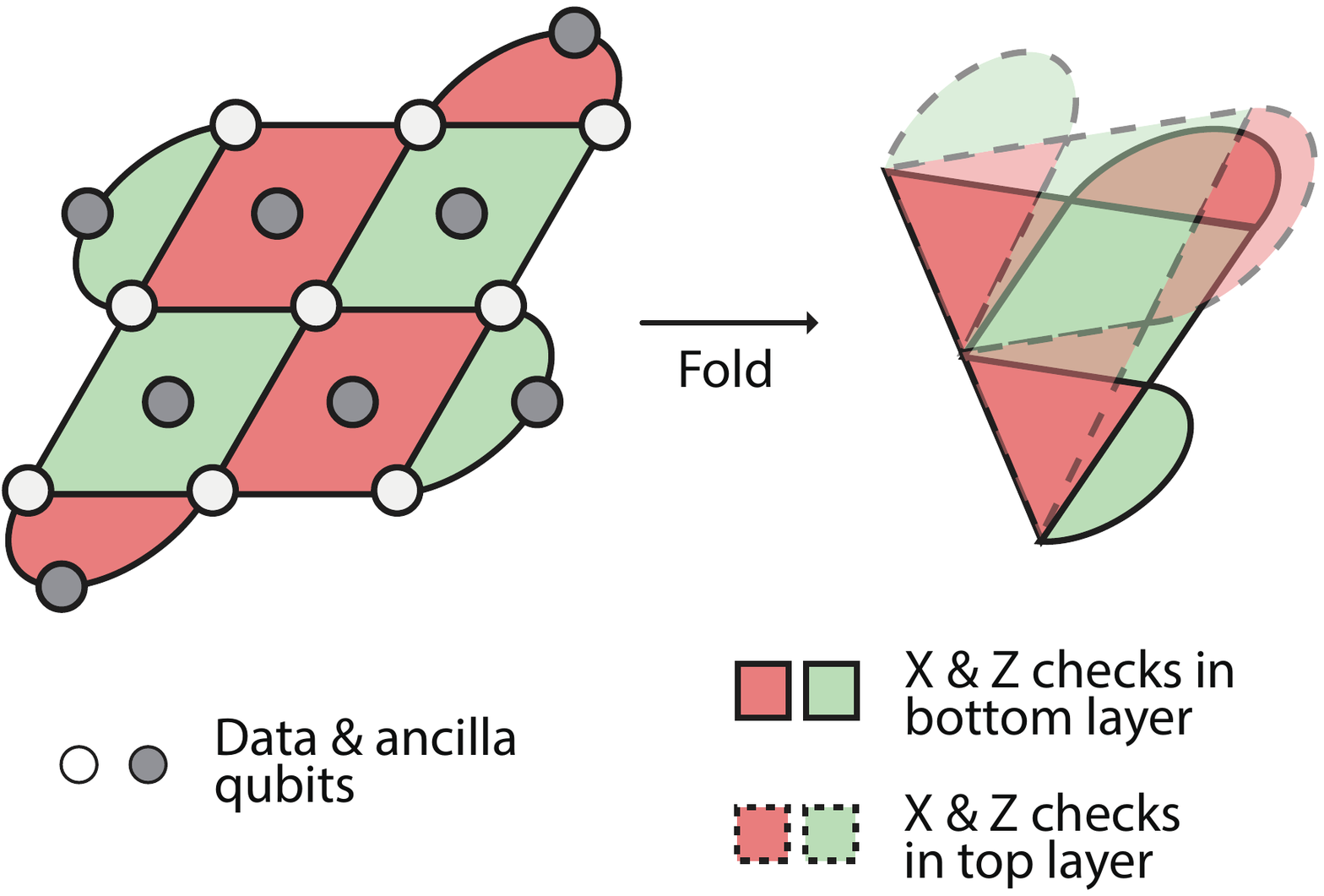}
    \caption{A distance 3 rotated surface code patch and the corresponding folded patch.}
    \label{fig:rotated_folded}
\end{figure}

Surface code \cite{bravyi1998quantum,kitaev2003fault} is considered as one of the most promising QEC codes for first-generation fault-tolerant quantum computers. If not specified, the word ``surface code' nowadays usually refers to the rotated surface code $\left[\left[d^2,1,d\right]\right]$, where $d$ is the distance of the code. As we will see, there are many variants of surface code. An example of a surface code patch with code distance 3 is shown at the left of \cref{fig:rotated_folded}. For completeness, the unrotated surface code is a $\left[\left[d^2+(d-1)^2,1,d\right]\right]$ CSS code.

In addition to the $d^2$ data qubits, each code patch requires $d^2-1$ ancilla qubits for stabiliser checks. One round of check is referred to as a code cycle, during which 4 layers of CNOT gates are applied in a specific sequence between the data and ancilla qubits to check $X$ and $Z$ stabilizers. The duration of a single cycle is referred to as the code cycle time.

The most commonly studied method for performing logical operations on surface code is called lattice surgery \cite{horsman2012surface}, based on merging and splitting the code patches, which has the benefit of requiring only 2D connectivity. In the standard lattice surgery setting, Hadamard gates are implemented by first applying transversal Hadamard then a 90 degrees patch rotation to reorient the code patch. The time taken for the transversal Hadamard is negligible compared to that of patch rotation. A common choice for patch rotation requires 1 ancillary patch and $3d$ code cycles \cite{litinski2019game}. Alternatively, if the space is limited, the rotation can also be done without ancilla in $1.5d$ cycles, but the code distance is halved \cite{fowler2018low}.

\begin{figure}[h]
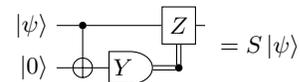

\begin{yquantgroup}
    \registers{
        qubit {} a;
        qubit {} b;
    }
    \circuit{
        init {$\ket{\psi}$} a;
        init {$\ket{0}$} b;

        cnot b|a;

        dmeter {$Y$} b;
        z a | b;
        discard b;
    }
    \equals[$=S\ket{\psi}$]
\end{yquantgroup}
\caption{$S$ gate implemented by Y basis measurement.}
\label{fig:tele_S}
\end{figure}

To implement logical $S$ gate via lattice surgery, catalyst state \cite{PhysRevA.86.032324}, twist defects \cite{PhysRevX.7.021029} and many other methods \cite{PRXQuantum.3.010331,gidney2024inplace} can be used. The most efficient method known \cite{gidney2024inplace} is based on logical $Y$ basis measurements, which requires 1 ancilla and $1.5d$ cycles. Here we will introduce the circuit in \cref{fig:tele_S} which operates on similar principles, where the Y basis measurement takes $d/2+2$ cycles \cite{gidney2024inplace}.

The logical CNOT gate can be implemented in $2d$ cycles using 1 ancilla \cite{Litinski2018latticesurgery}. A related gate is SWAP, decomposing SWAP into 3 CNOT will cost 1 ancilla and $6d$ cycles. SWAP gate can also be implemented by patch movement, which takes 2 ancillae and $2d$ cycles (assuming the two patches are placed diagonally).

\textbf{Folded surface code:} as its name suggests, the folded surface code is obtained by folding a regular surface code patch along its diagonal \cite{PhysRevA.94.042316}. We note that folding either an unrotated or a rotated surface code results in what is generally referred to as a folded surface code patch; however, these two types of patches differ in structure. In this work, we focus on the folded rotated surface code, with an example patch shown on the right of \cref{fig:rotated_folded}. \black{In this work, the fold is taken to lie along the logical $Z$ operator by default. Consequently, the ``hypotenuse" of the folded surface code patch supports a logical $Z$ operator, while the two ``legs" support both $X$ and $Z$ operators.} As we will elaborate in the following sections, recent developments \cite{McEwen2023relaxinghardware,chen2024transversal} allow all the Clifford generators $\{H,S,CNOT\}$ to be implemented transversally on folded rotated surface code. In contrast to the $O(d)$ runtime of lattice-surgery-based Clifford gates, the runtime of transversal Clifford gates is independent of $d$. In practice, for the code distances of interests ($20\lesssim d\lesssim30$), the spacetime volume of transversal Clifford gates is typically one to two orders of magnitude smaller than that of their lattice surgery counterparts.

\textbf{Yoked surface code} \cite{gidney2025yoked} is a concatenation of rotated surface codes (inner code) and quantum parity check codes (outer code). While the inner surface code patches have their own stabiliser checks, the outer code checks the $X$ and $Z$ parity of the encoded surface code patches. By combining this information with the soft information from inner codes, the overall effect of this concatenation is that the yoked surface code patches can achieve same logical error rate with smaller code distance compared to unyoked surface code. This means that, although the parity checks incur extra space overhead, the overall storage efficiency (logical qubits per physical qubit) can increase by up to a factor of approximately three. There are two types of yoked surface code introduced in ref.~\cite{gidney2025yoked}, called 1D and 2D yoked surface code, using 
$\left[\left[n,n-2,2\right]\right]$ codes and $\left[\left[n^2,n^2-4n+2,4\right]\right]$ codes as outer code respectively. 

\subsection{Standard looped pipeline architecture}

\begin{figure}[h]
    \centering
    \subfloat[Implementation of one 2D qubit layer.]{
    \includegraphics[width=0.8\linewidth]{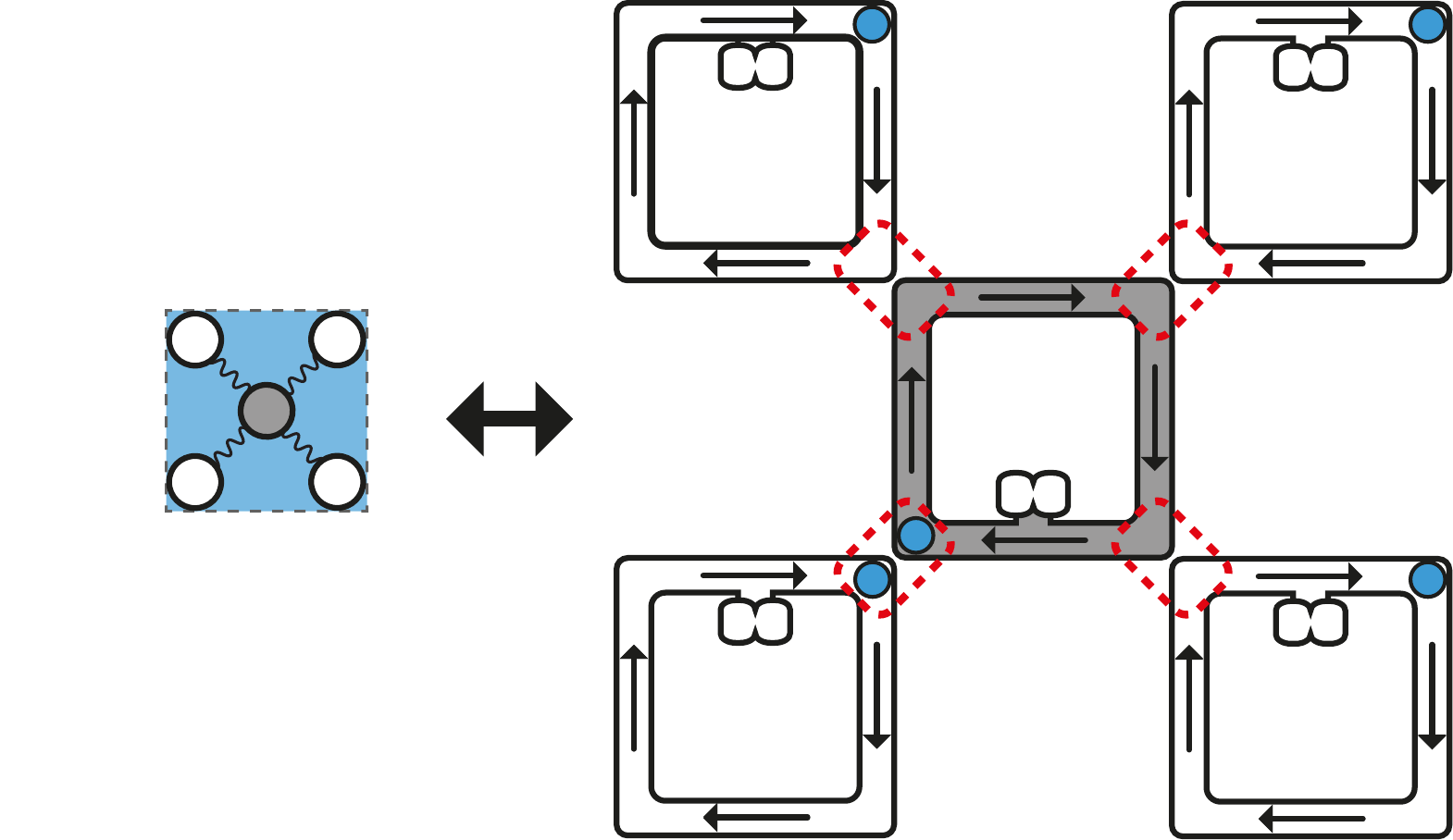}
    \label{fig:loop_a}
    }\\
    \subfloat[Implementation of four 2D qubit layers in a stack.]{
    \includegraphics[width=0.8\linewidth]{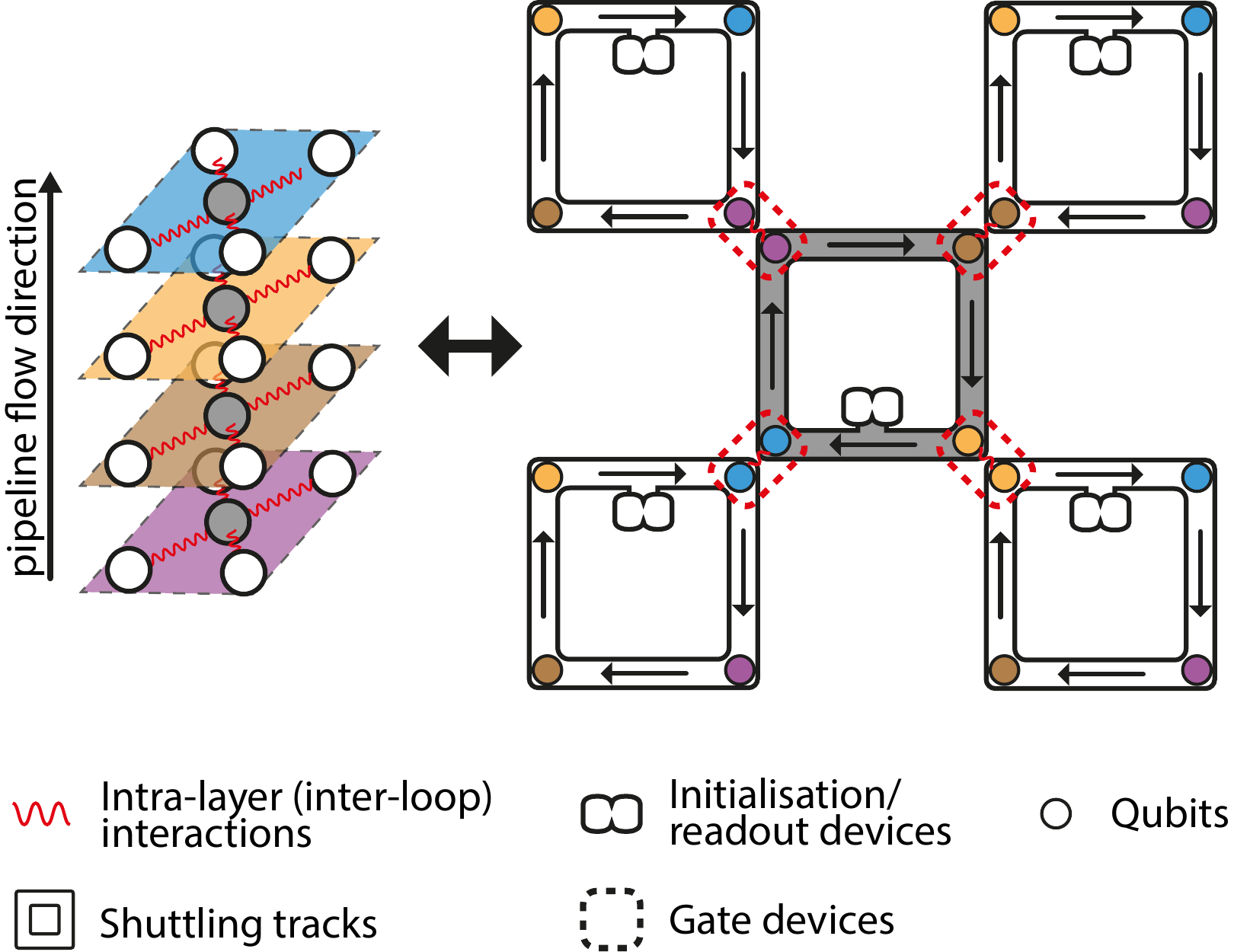}
    \label{fig:loop_b}
    }
    \caption{The looped pipeline architecture. By placing more qubits in each loop via pipelining, we can `stack' the patches.}
\end{figure}

For many hardware platforms, qubit shuttling is often required to increase inter-qubit spacing, thereby reducing crosstalk, and in some cases to accommodate control hardware. While this introduces engineering challenges, it also creates opportunities for more advanced architectures beyond simple nearest-neighbour connectivity. One such scalable architecture that leverages qubit shuttling is the looped-pipeline architecture \cite{PRXQuantum.4.020345}. The basic idea of this architecture is to use shuttling loops; an example implementation with five qubits arranged in a square-grid layout is shown in \cref{fig:loop_a}. The advantage of this implementation emerges as more qubits are added to each loop. With careful synchronization and pipelining, additional qubit patches can be hosted while preserving the same connectivity and without incurring extra space overhead, as shown in \cref{fig:loop_b}. Conceptually, this corresponds to a \textit{virtual stack} of qubit patches while remaining a strictly 2D architecture. Further enabling the interactions between qubits within the same loop (intra-loop interactions) is then equivalent to coupling qubits across different layers of the stack (inter-layer interactions), effectively realizing transversal operations between different qubit patches.

\begin{figure}
    \centering
    \includegraphics[width=0.9\linewidth]{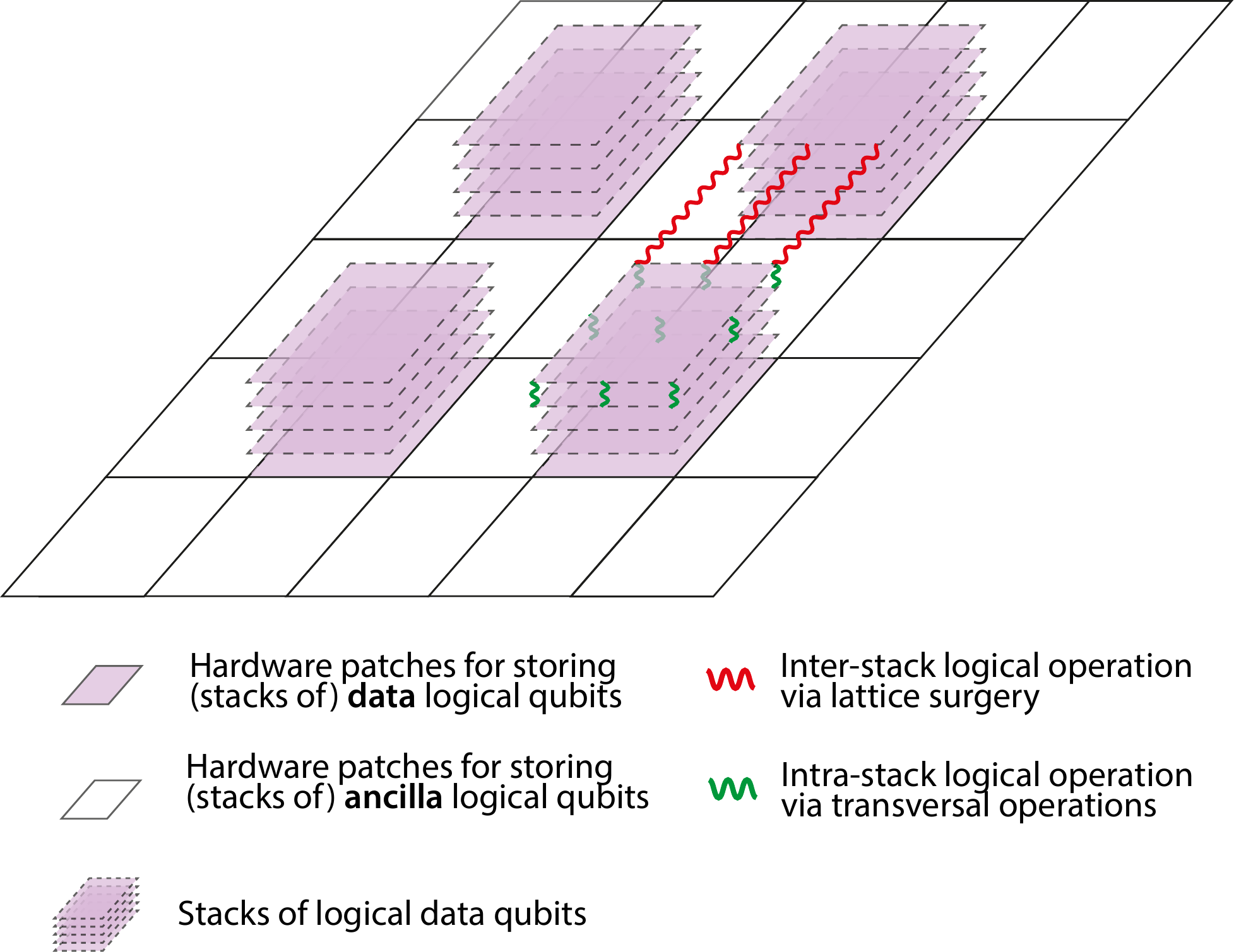}
    \caption{The hallway layout for stacks of logical qubits. Both data (purple) and ancilla (white) code patches are stored in stacks. While some intra-stack operations are transversal, inter-stack operations still require lattice surgery.}
    \label{fig:stacks_hallway}
\end{figure}

\begin{figure*}
    \centering
    \includegraphics[width=0.85\linewidth]{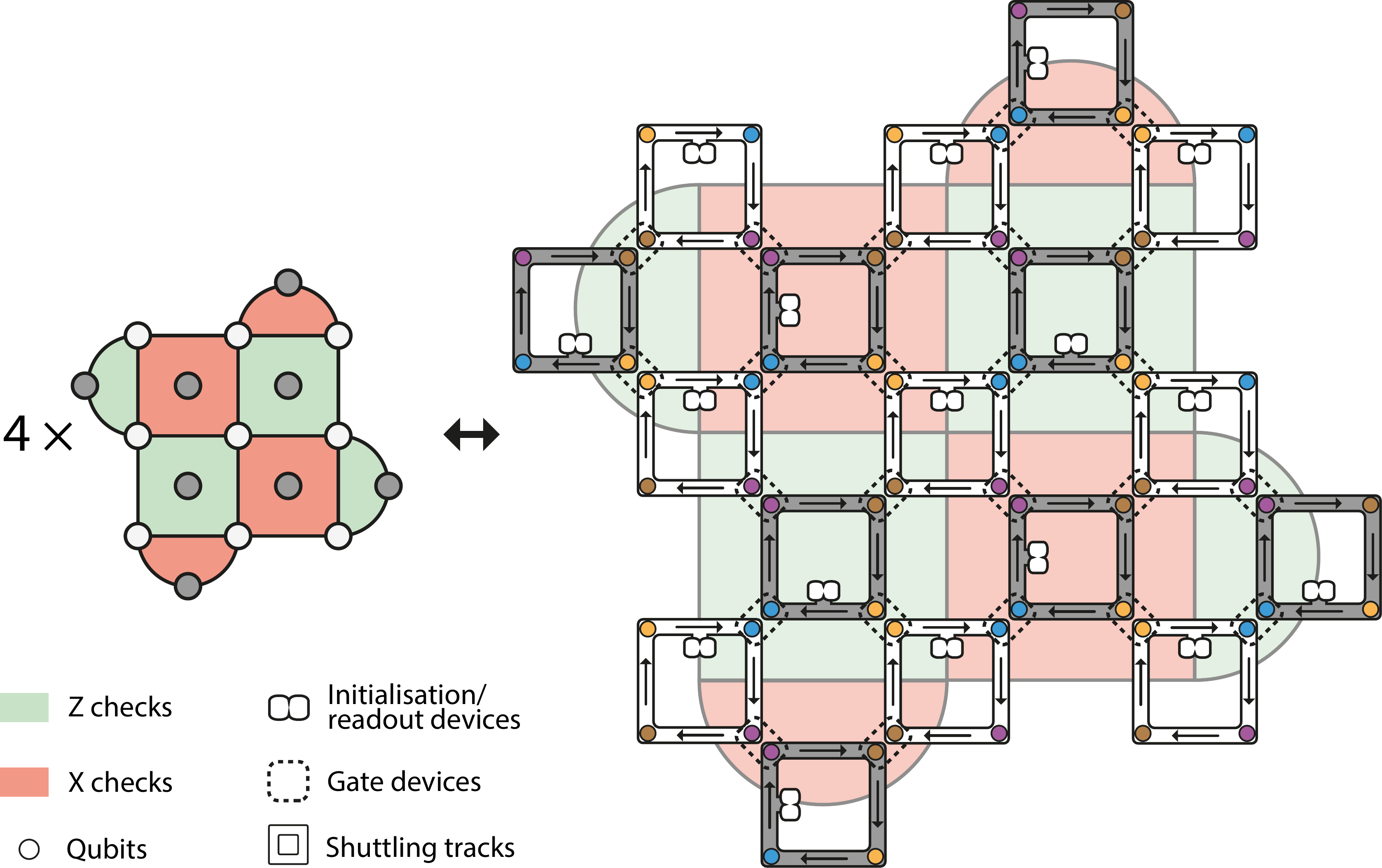}
    \caption{A stack of distance-3 surface code patches implemented using looped pipeline architecture.}
    \label{fig:surface_stack}
\end{figure*}

It is also possible to stack surface code patches and perform stabilizer checks within the looped pipeline architectures. An example of a stack of distance-3 surface code patches is shown in \cref{fig:surface_stack}. However, as we will see later, a taller stack implies more qubits per loop, which can lead to congestion and introduce waiting times during stabilizer checks, resulting in a longer code cycle time. Additional readout devices may be required in each loop to shorten this pipelining delay.

In contrast to standard lattice surgery, the availability of transversal operations can greatly improve the speed of several logical gates. If intra-loop interactions are enabled, CNOT gates between two code patches in the same stack can be implemented transversally. This transversal CNOT can also help with the implementation of logical $S$ gate using the circuit in \cref{fig:tele_S}, but the Y basis measurement still takes $O(d)$ code cycles. Moreover, because our loop pipeline architecture permits only short-range shuttling within the shuttling loop, the Hadamard gate, or more precisely, the patch rotation required for a logical Hadamard, still relies on lattice surgery. The operations (e.g. CNOT) between different stacks also require lattice surgery. We also note that, the ability to perform intra-loop interaction was assumed in ref.~\cite{PRXQuantum.4.020345}, but without detailing an explicit implementation. In the next section, we will present a possible scheme for implementing intra-loop interaction.

\begin{figure*}
    \centering
    \includegraphics[width=0.85\linewidth]{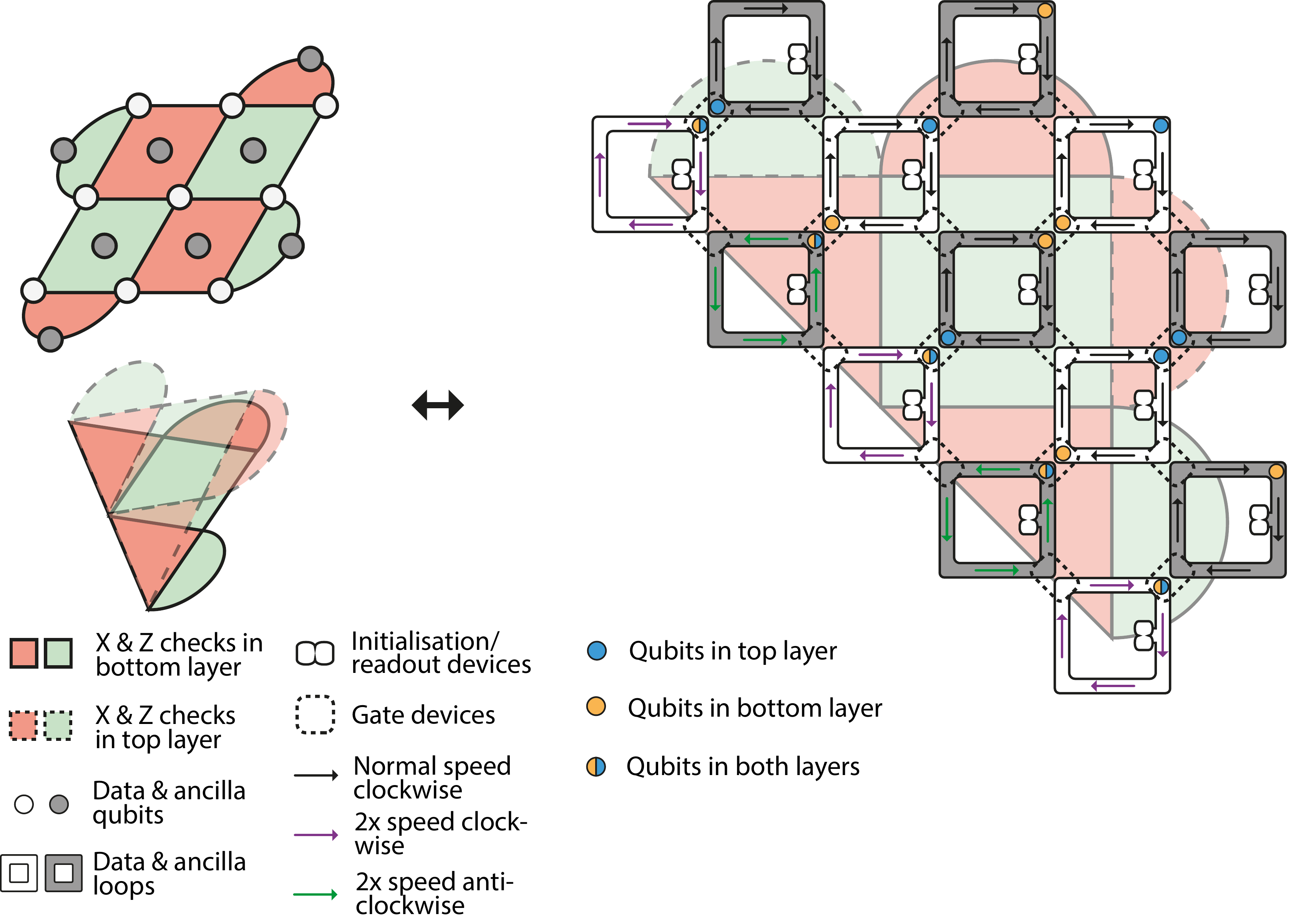}
    \caption{Folded surface code implemented by looped pipeline. To connect the top and bottom layers, the qubits along the diagonal are shuttled two times faster (indicated by green and purple arrows).}
    \label{fig:fold1}
\end{figure*}

At the large scale, the layout scheme for stacks of code patches proposed in ref.~\cite{PRXQuantum.4.020345} is the quasi 3D analogue of the ``hallway'' layout, which is widely used in architectures with nearest-neighbour connectivity. An illustration is shown in \cref{fig:stacks_hallway}. This generic layout is capable of handling a wide range of computational scenarios and is especially suitable for highly parallel operations. However, it does not exploit the full potential of the virtual stacks. We will show in \cref{Sec:architecture} that the hallway layout is merely a special case of a more general and flexible architecture.

\section{intra-loop interactions}
\label{Sec:rearrange}

We have shown how stacks of two-dimensional code patches can be implemented using the looped pipeline architecture, and in later sections we extend this framework to folded surface codes. Interactions between distinct code layers within a stack map onto interactions between the corresponding physical qubits within a single loop. We now describe methods for implementing such intra-loop interactions. In particular, we focus on implementing physical two-qubit gates between arbitrary pairs of qubits within a loop, which correspond to transversal gates between any two layers of code patches in the same stack. In addition, we discuss a more complex procedure in \cref{appx:rearrangement} for rearranging qubits within a loop into an arbitrary ordering. This corresponds to a reordering of code layers within the stack and can facilitate the efficient implementation of multiple two-qubit gates. Throughout, we assume that qubits can be shuttled both clockwise and anticlockwise, and can be held stationary when required.

\begin{figure*}
    \centering
    \includegraphics[trim={2cm 20cm 2cm 9cm},clip,width=0.95\linewidth]{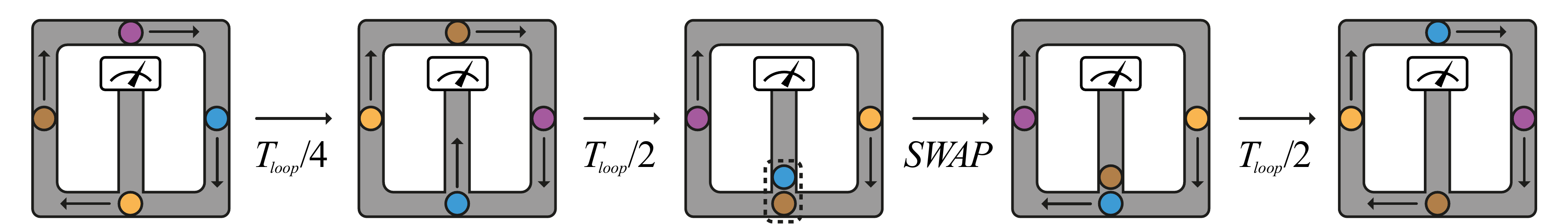}
    \caption{Example of a SWAP operation between the blue and brown qubits. This corresponds to a worst-case scenario in which the qubits initially occupy the positions indicated at the beginning of the flow chart, requiring a shuttling time of \(\tfrac{5}{4} T_{loop}\). The SWAP gate may be replaced by any two-qubit gate to generalize this protocol. At the end of the branch track is the measurement/initialisation device.}
    \label{fig:SWAP}
\end{figure*}

In order to see how the intra-loop interactions are realised, we will zoom into the detailed structure of the loops as shown in \cref{fig:SWAP}. We see that there is a junction in each loop branching out to a port of qubit parking space, operating in the `last in, first out' manner. Gate devices are placed near the junction to enable intra-loop qubit interactions by bringing qubits together using the junction. A measurement/initialisation device is placed at the end of the port, and additional such devices can be added to enable parallel measurement/initialisation. We note that, for platforms like silicon qubits, adding junctions to shuttling tracks can be challenging, therefore such an one-junction structure is one of the approaches that place minimal additional demands on hardware engineering. From here on, we will assume all loops shown in the previous figures come with such a structure.

Executing a physical two-qubit gate (using SWAP as an example) between qubit pair $(a,b)$ in the new loop requires 4 steps, as also illustrated in \cref{fig:SWAP}:

\begin{enumerate}
    \item Shuttle qubit $a$ into the port. Without loss of generality, we assume that qubit $a$ is closer to the port entrance.
    \item Shuttle qubit $b$ into the port.
    \item Apply SWAP between qubit $a$ and $b$. During this process, all other qubits in the loop are held fixed.
    \item Shuttle qubit $b$ out of the port to qubit $a$'s original position.
\end{enumerate}

We use $T_{loop}$ to denote the time required to shuttle a qubit through one complete lap around the loop. Then the shuttling time required to perform a SWAP operation between any two qubits is at most \(\tfrac{5}{4} T_{loop}\) as shown in \cref{fig:SWAP}. Using the same shuttling scheme, we can implement any two-qubit gate between an arbitrary pair of qubits within the same loop by replacing the SWAP operation with the target two-qubit gate. The upper bound on the time required for inter-loop interactions between arbitrary qubits obtained here will form the basis for the timing requirements of transversal gates discussed later.

An alternative protocol for implementing a SWAP gate between arbitrary qubit pairs is through a physical SWAP. In \cref{fig:SWAP}, at some point during the first `$T_{loop}/2$', the blue qubit is in the port while the purple qubit has already passed the port's entry. At this stage, we can shuttle the blue qubit out and then shuttle the brown qubit in. After another `$T_{loop}/2$', we shuttle the brown qubit out, completing the protocol. However, this results in uneven spacing between the qubits, which may disrupt the synchronization required for stabilizer checks. The blue qubit needs to be shuttled into the port (or slowed down) for a short time to restore the synchronization, which can incur extra time overhead.

\section{looped pipeline with folded surface code}
\label{Sec:folded}
\subsection{General Implementation}
As illustrated in \cref{fig:fold1}, the looped pipeline architecture can also be used to implement the folded surface code. The folded surface code occupies roughly half as many loops as the standard surface code, with the two folded halves realised as two qubit layers in the looped pipeline—each loop containing two qubits, except along the diagonal and at the ancilla loops on the boundary. Along the diagonal (the crease of the fold), each loop hosts only a single qubit, which must be shuttled at twice the speed of qubits in off-diagonal loops in order to connect the top and bottom layers. For simplicity, we assume that all loops can operate at either the normal or double shuttling speed, allowing the triangular code patches to be placed anywhere on the grid.

A key advantage of the folded surface code is the transversality of Clifford gates. The CNOT gate is natively transversal for the rotated surface code and therefore also for the folded surface code (and, in fact, for all CSS codes acting between two code blocks). For the folded rotated surface code used here, the logical $S$ gate can be implemented transversally during the middle of a stabilizer measurement round. The detailed scheme, following ref.~\cite{chen2024transversal}, is as follows:
\begin{enumerate}
    \item carry out the first two layers of CNOT of a syndrome measurement as usual,
    \item apply physical $S$ and $S^\dagger$ gates alternately to the diagonal qubits, along with CZ gates between folded pairs,
    \item complete the syndrome measurement with the remaining two layers of CNOT gates.
\end{enumerate}

\begin{figure}
    \centering
    \includegraphics[width=0.8\linewidth]{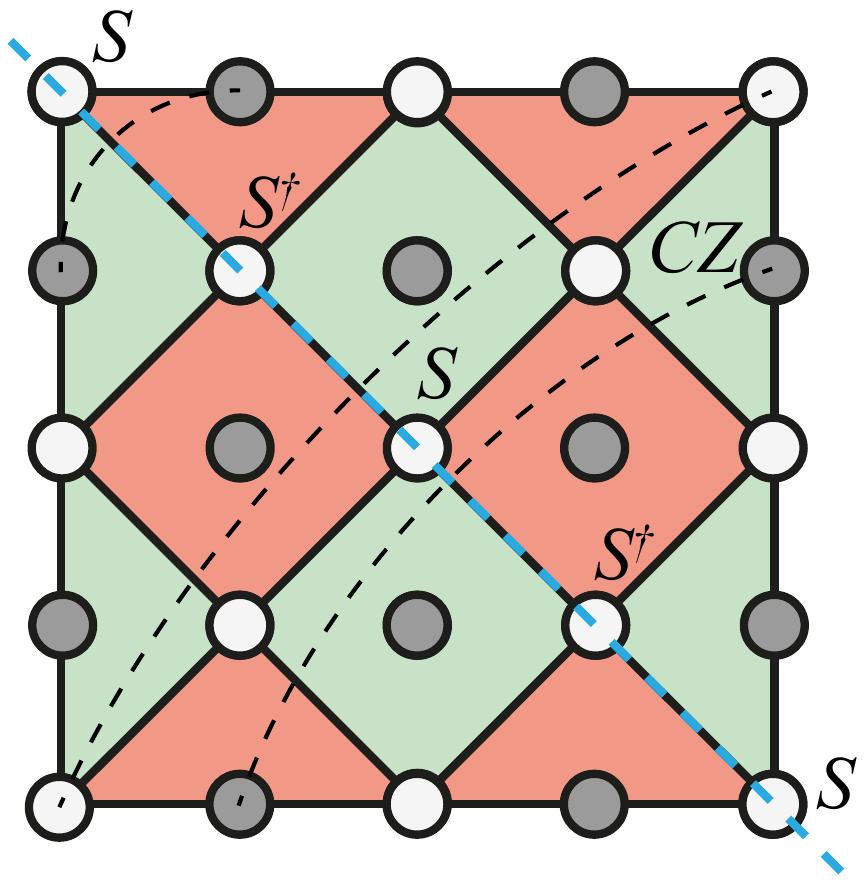}
    \caption{\black{Transversal implementation of the \(S\) gate for the unrotated surface code. The $S$/$S^{\dagger}$ gates are applied to qubits along the diagonal (blue dashed line), while $CZ$ gates are applied between all folded pairs. Three example folded pairs are are shown (black dashed lines).}}
    \label{fig:Gate}
\end{figure}

In essence, this protocol relies on the fact that, after the first two layers of CNOT gates, the rotated surface-code patch is transformed into an unrotated surface-code patch \cite{McEwen2023relaxinghardware}. The transversal \(S\) scheme for the unrotated surface code \cite{PhysRevA.94.042316} can then applied, \black{as illustrated in \cref{fig:Gate}}. Finally, the remaining two layers of CNOT gates transform the patch back into the rotated surface code. The logical Hadamard gate can be implemented in a similar manner, except that the operations in the second step differ: Hadamard gates are applied transversally to all data and ancilla qubits (excluding the two-body stabilizer ancillae at the boundary), followed by SWAP gates between folded pairs. \black{Furthermore, recall that in standard lattice surgery, a logical Hadamard operation consists of a transversal Hadamard followed by a patch rotation. Therefore, if we apply our ``mid-cycle” transversal Hadamard protocol and subsequently perform the lattice-surgery Hadamard without the patch rotation (followed by one round of syndrome extraction), the logical operator supported along the fold diagonal is transformed from $Z$ to $X$ (or from $X$ to $Z$). This transformation is equivalent to a 90 degrees patch rotation in the regular rotated surface code, which we refer to as diagonal switching.}

We first go through the scheme for implementing a single round of stabiliser checks for a single folded surface code patch in our architecture, with reference to \cref{fig:fold1}. The order of CNOT gates for a rotated surface code patch (top left of \cref{fig:fold1}) is
$$
\text{$X$ checks}
\hspace{0.3cm}
\begin{matrix}
1 & \rightarrow & 2\\
 & \swarrow & \\
3 & \rightarrow & 4
\end{matrix}
\hspace{2cm}
\begin{matrix}
1 &  & 3\\
\downarrow & \nearrow & \downarrow\\
2 & & 4
\end{matrix}
\hspace{0.3cm}
\text{$Z$ checks}
$$

This order is kept even when the code patch is folded. For example, on the right of \cref{fig:fold1}, for the ancilla loop at the center of the green square (which is a $Z$ check for both layers), the order of CNOT gates is
$$
\text{Top qubit}
\hspace{0.3cm}
\begin{matrix}
1 & \rightarrow & 2\\
 & \swarrow & \\
3 & \rightarrow & 4
\end{matrix}
\hspace{2cm}
\begin{matrix}
1 &  & 3\\
\downarrow & \nearrow & \downarrow\\
2 & & 4
\end{matrix}
\hspace{0.3cm}
\text{Bottom qubit}
$$

Assume, without loss of generality, that at the start of the stabilizer cycle the yellow qubits already occupy the positions required for the first layer of CNOT gates in the check circuit. The first two layers of CNOT require $\frac{5}{4}T_{loop}$ shuttling time plus $2T_{2q}$ for the CNOT gates for all loops to finish, where $T_{2q}$ denotes the two-qubit gate time. Once this is done, the qubits are at the exact locations shown in \cref{fig:fold1}. The remaining two layers of CNOT for the checks take another $\frac{5}{4}T_{loop}+2T_{2q}$. Assuming the measurement time is $T_{meas}$ and 3 measurement devices per loop (reasonable in our later example of silicon qubits), completing the measurement would take $T_{meas}$ and an extra $\frac{7}{8}T_{loop}$ shuttling time to enter the port. Taking into account the Hadamard gate time $T_{1q}$ at the beginning and the end of the checks, we can conclude that $T_{cyc}(n=2)=\frac{27}{8}T_{loop}+2T_{1q}+4T_{2q}+T_{meas}$ (see \cref{appx:stabiliser} for more details). This is for the case where there are $2$ qubits per loop ($n = 2$), i.e. single folded surface code patch. Adding more qubits per loop can introduce delays and thus increase the code cycle time, but careful pipelining the operations on different qubits can mitigate this overhead. As a result, the approach is highly dependent on the exact time required for gates, shuttling and measurement. We will discuss this in our example later in \cref{sec:silicon_implem}. 

\begin{figure*}
    \centering
    \includegraphics[width=\linewidth]{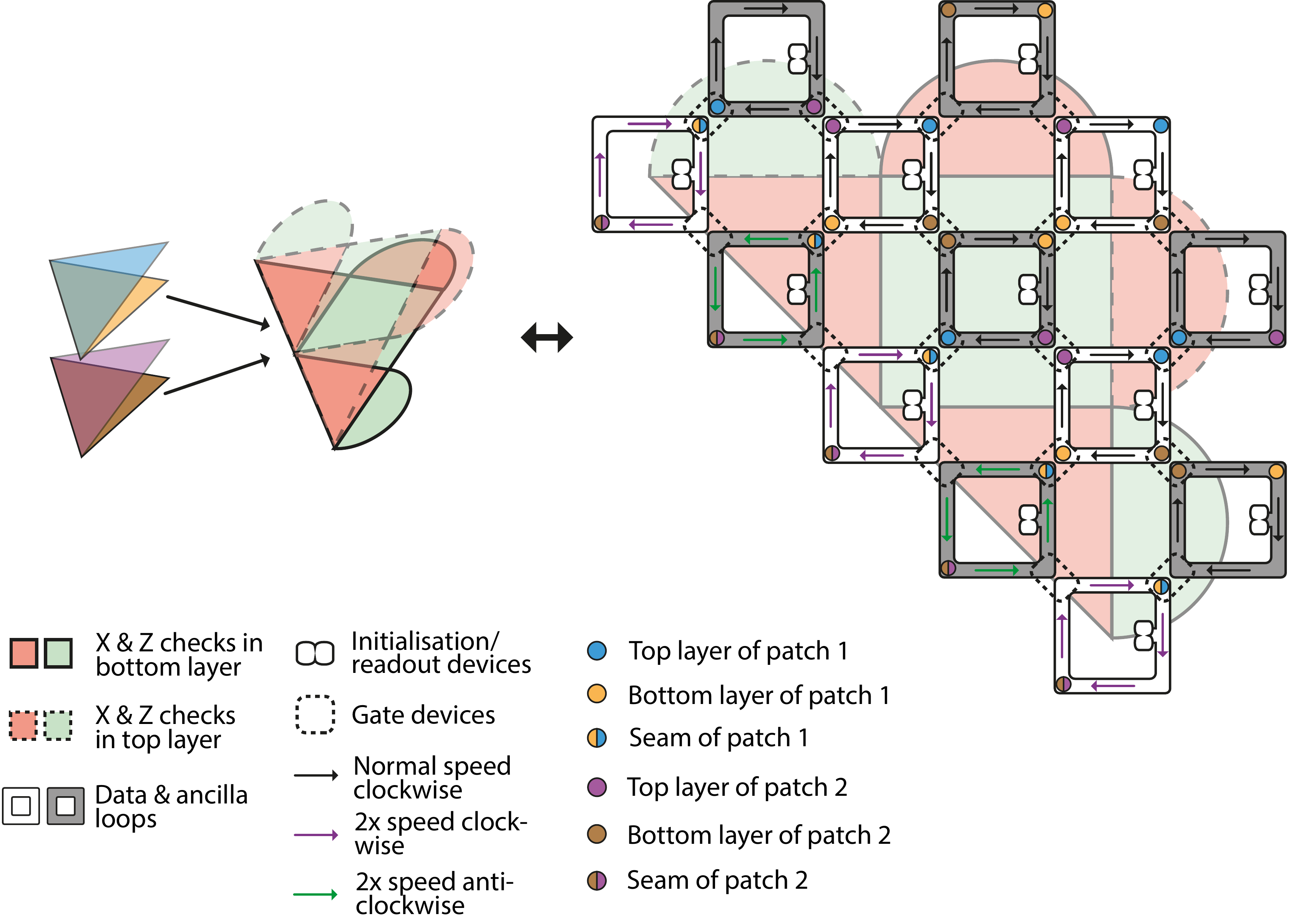}
    \caption{Two layers of folded surface code implemented by looped pipeline.}
    \label{fig:fold2}
\end{figure*}

We now modify this protocol to implement the transversal logical $S$ gate for the folded surface code in our architecture, with reference to the single code patch case in \cref{fig:fold1} for simplicity. Right after the first two layers of CNOT, we apply $S$ or $S^\dagger$ to the diagonal qubits and use the intra-loop interaction protocol in \cref{Sec:rearrange} to apply $CZ$ to the off-diagonal pairs, and this takes $\frac{9}{8}T_{loop}+T_{2q}$. Right after the $CZ$, the yellow qubit (without loss of generality) is at the port's entry, we then reversely shuttle (i.e. in the direction opposite to the arrows in \cref{fig:fold1}) the qubits in both diagonal and off-diagonal loops for $\frac{1}{8}T_{loop}$, so that the yellow qubit is at the correct place for the next layer of CNOT. Then we proceed with the remaining two layers of CNOT. The overall time overhead for logical $S$ gate is thus 
\begin{equation}
\label{eqn:T_S}
    T_S=T_{cyc}+\frac{5}{4}T_{loop}+T_{2q}.
\end{equation}
 
Using a similar procedure, the logical $H$ can be applied in 
\begin{equation}
\label{eqn:T_H}
T_H=T_{cyc}+\frac{5}{4}T_{loop}+T_{1q}+T_{2q},
\end{equation}
\black{and the diagonal switching takes $T_{ds}=T_{H}+T_{1q}+T_{cyc}$.}

Finally, we present a scheme for implementing transversal SWAP/CNOT gates between two folded surface code patches within the same stack, as illustrated in \cref{fig:fold2}. The extension to a larger number of patches is straightforward. To realize a transversal gate within a single stack, we apply the intra-loop interaction protocol in \cref{Sec:rearrange} twice: once for the top layers and once for the bottom layers of the folded patches. The worst-case time required for this operation is
\begin{equation}
\label{eqn:T_CNOT}
    T_\mathrm{CNOT}(n)=\left(\frac{9}{4}-\frac{7}{2n}\right)T_{loop}+2T_{2q}
\end{equation}
for $n$ qubits in a loop. And $T_\mathrm{SWAP}(n)=T_\mathrm{CNOT}(n)$.

\subsection{Example: Silicon Spin Qubits}\label{sec:silicon_implem}
In the case of silicon spin qubits, ref.~\cite{de2025high} have demonstrated high-fidelity single-electron spin shuttling at speeds up to \(64\,m/s\), while simulation results in ref.~\cite{PhysRevB.102.125406} indicate that high fidelity shuttling can be achieved with a speed up to \(100\,m/s\). When considering folded surface code, we therefore assume a shuttling speed of \(50\,m/s\) in the off-diagonal loops (and a shuttling speed of \(100\,m/s\) in the diagonal loops). For a loop perimeter of \(L_{loop}=20\,\mu m\), this yields \(T_{loop} = 400\,ns\) for the off-diagonal loops (which is the majority of the loops). The time needed for shuttling qubit around the diagonal loops is thus $T_{loop}/2$. For gate operation time and measurement time, we will mostly follows the number quoted in ref.~\cite{PRXQuantum.4.020345} with single-qubit gate time $T_{1q} \sim 200\, ns$, two-qubit gate time $T_{2q} \sim  100\, ns$, measurement time $T_{meas}\sim 1\ \mu s$, and with 3 measurement devices per loop. We note that in ref.~\cite{PRXQuantum.4.020345} the quoted single qubit gate time $\sim 25\ ns$ assumes the use of micromagnets. However, since micromagnets can introduce additional engineering challenges, we opt for an architecture without them, leading to our assumption of $T_{1q}=200\ ns$ (which requires improvements upon the MHz-scale Rabi frequency achieved in \cite{PhysRevApplied.21.014044,unseld2025baseband}).

Using arguments in the last section, the stabiliser code cycle time for implementing one folded surface code in the stack ($n=2$ qubits per loop), $T_{cyc}(n=2)$, evaluates to $3.15\ \mu s$ for spin qubits. For multiple code patches, we focus on the case of 16 qubits per loop ($n=16$) for later use in magic state factory. With three measurement devices, if we wait for all qubits to be measured before initiating the next code cycle, the cycle time is approximately $8.15\ \mu s$. Alternatively, say the check circuit begins at $t=0$, then at $t=3.15\ \mu s$ the first three qubits that complete their measurements in cycle 1 can enter the next code cycle. At $t=4.15\ \mu s$, three more qubits enter code cycle 2, lagging the first group of three qubits in cycle 2 by $1\ \mu s$. By $t=5.3\ \mu s$, nine qubits have entered cycle 2 and three of them are ready for measurement 2. However, seven qubits have not yet completed cycle 1, and the nine qubits that have entered cycle 2 must therefore wait. At $t=7.15 \ \mu s$, only a single qubit remains to be measured in cycle 1, at which point two qubits can begin their second-round measurements using the two available measurement devices. By $t=8.15 \ \mu s$, all qubits completed cycle 1, and two qubits can enter cycle 3. For these two qubits, the average code cycle time up to this point is $4.075 \ \mu s$. Continuing this process, the average code cycle time converges to $\sim 5.3\ \mu s$, which is just $16/3\times 1\mu s$. This result is reasonable, as we assume that other operations, such as CNOT gates, can be executed while awaiting measurement outcomes; consequently, the code cycle time in this case is limited primarily by the measurement duration. To account for potential congestion, we round this value up and take $T_{cyc}^*(16)\sim 6\ \mu s$ in the following discussion.

For $n=16$ case, using \cref{eqn:T_S,eqn:T_H}, logical $S$ requires 6.6 $\mu s$ and logical $H$ requires 6.8 $\mu s$. And one can just use $T_S\sim T_H\sim T_{cyc}$ for estimation. Following \cref{eqn:T_CNOT}, transversal CNOT gate within the same stack requires $\sim  1\ \mu s$ for silicon spin qubits. The CNOT gates between stacks, however, still require lattice surgery.

\section{Magic State Factory}
\label{Sec:magic}

With all the ingredients ready, we now turn to the magic state factories. We implement the 8T-to-CCZ factory in \cite{fazio2025low}, which is optimised for transversal CNOT and $S$. The circuit diagram is shown in \cref{fig:CCZ_folded}, with time slices 1 to 7. One round of stabiliser checks is performed at each time slice. To enable transversal gates, we need to put the qubits in the same virtual stack. The circuit in \cref{fig:CCZ_folded} operates on 8 qubits, implying that each loop need to host up to 16 qubits.

The input $T$ states are generated via magic state cultivation \cite{gidney2024magic}. We assume an output logical error rate of $10^{-7}$ for $T$ cultivation, which in turn gives a $CCZ$ error rate of $28\times(10^{-7})^2=2.8\times10^{-13}$. Assuming a code distance $d=25$, which is sufficient for large scale applications such as factoring 2048-bit RSA integers \cite{gidney2025factor}, cultivating the 8 input $T$ states on 8 logical qubits requires 22 code cycles. Using the results of \cref{Sec:folded}, we conclude that the overall runtime using our architecture for silicon spin qubits is 216 $\mu s$ for $n=16$. Further details of this runtime calculation are provided in \cref{appx:CCZ}.

\begin{figure*}
    \centering

\begin{tikzpicture}
\begin{yquant*}[operator/minimum width=0pt, operator/separation=1mm]
qubit {$q_{\idx}:\ket{T}$} a[8];

cnot a[0] | a[1];
cnot a[3] | a[2];

[red, label=1] barrier (a);
cnot a[2] | a[0];
cnot a[1] | a[3];

[red, label=2] barrier (a);
cnot a[0] | a[1];
cnot a[3] | a[2];

[red, label=3] barrier (a);
cnot a[4] | a[0];
cnot a[5] | a[1];
cnot a[6] | a[2];
cnot a[7] | a[3];

[red, label=4] barrier (a);
dmeter {$Z$} a[4-7];

box {$S$} a[0] | a[4];
discard a[4];
box {$S$} a[1] | a[5];
discard a[5];
box {$S$} a[2] | a[6];
discard a[6];
box {$S$} a[3] | a[7];
discard a[7];

[red, label=5] barrier (a);
align a[0], a[3];
cnot a[0] | a[1];
cnot a[2] | a[3];

[red, label=6] barrier (a);
cnot a[1] | a[3];

[red, label=7] barrier (a);
X a[0-2];

hspace {2mm} a[0];
output {$\ket{CCZ}$} (a[0,1,2]);
output {$\bra{+}$} a[3];

\end{yquant*}
\end{tikzpicture}
\caption{Circuit diagram of the 8T-to-CCZ factory optimized for transversal CNOTs from ref.~\cite{fazio2025low}. The T states are produced by magic state cultivation. One round of stabiliser checks is performed at every time slice.}
\label{fig:CCZ_folded}
\end{figure*}
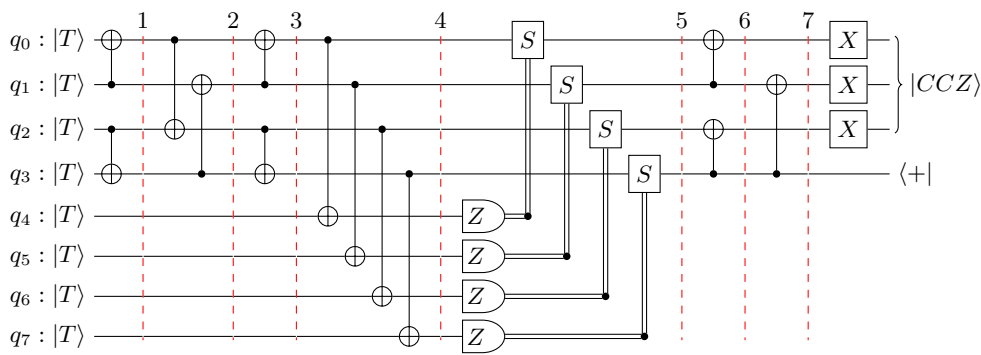

\begin{figure*}
    \centering
\begin{tikzpicture}
\begin{yquant*}[operator/minimum width=0pt, operator/separation=1.5mm]
qubit {} a[12];
init {$q_{\idx}:\ket{T}$} a[0-7];
init {$q_{\protect\the\numexpr\idx+8}:\ket{0}$} a[8-11];

cnot a[0] | a[1];
cnot a[3] | a[2];

[red, label=1] barrier (a[0-11]);
cnot a[2] | a[0];
cnot a[1] | a[3];

[red, label=2] barrier (a[0-11]);
cnot a[0] | a[1];
cnot a[3] | a[2];

[red, label=3] barrier (a[0-11]);
cnot a[4] | a[0];
cnot a[5] | a[1];
cnot a[6] | a[2];
cnot a[7] | a[3];

[red, label=4] barrier (a[0-11]);
dmeter {$Z$} a[4-7];
hspace {1cm} a[8-11];

hspace {10.8mm} a[0];
[name=d0]
zz a[0];
\draw (d0) |- + (0,-2.3);
cnot a[8]|a[4];
discard a[4];

hspace {14.9mm} a[1];
[name=d1]
zz a[1];
\draw (d1) |- + (0,-2.3);
cnot a[9]|a[5];
discard a[5];

hspace {19mm} a[2];
[name=d2]
zz a[2];
\draw (d2) |- + (0,-2.2);
cnot a[10]|a[6];
discard a[6];

hspace {23.1mm} a[3];
[name=d3]
zz a[3];
\draw (d3) |- + (0,-2.2);
cnot a[11]|a[7];
discard a[7];

[red, label=5] barrier (a[0-11]);
dmeter {$Y$} a[8-11];
Z a[0] | a[8];
discard a[8];
Z a[1] | a[9];
discard a[9];
Z a[2] | a[10];
discard a[10];
Z a[3] | a[11];
discard a[11];

[red, label=6] barrier (a[0-11]);
align a[0], a[3];
cnot a[0] | a[1];
cnot a[2] | a[3];

[red, label=7] barrier (a[0-11]);
cnot a[1] | a[3];

[red, label=8] barrier (a[0-11]);
X a[0-2];

hspace {2mm} a[0];
output {$\ket{CCZ}$} (a[0,1,2]);
output {$\bra{+}$} a[3];

\end{yquant*}
\end{tikzpicture}
\caption{Circuit diagram of the 8T-to-CCZ factory when a transversal $S$ gate is not available. The T states are produced by magic state cultivation. One round of stabiliser checks is performed at every red time slice.}
\label{fig:CCZ}
\end{figure*}
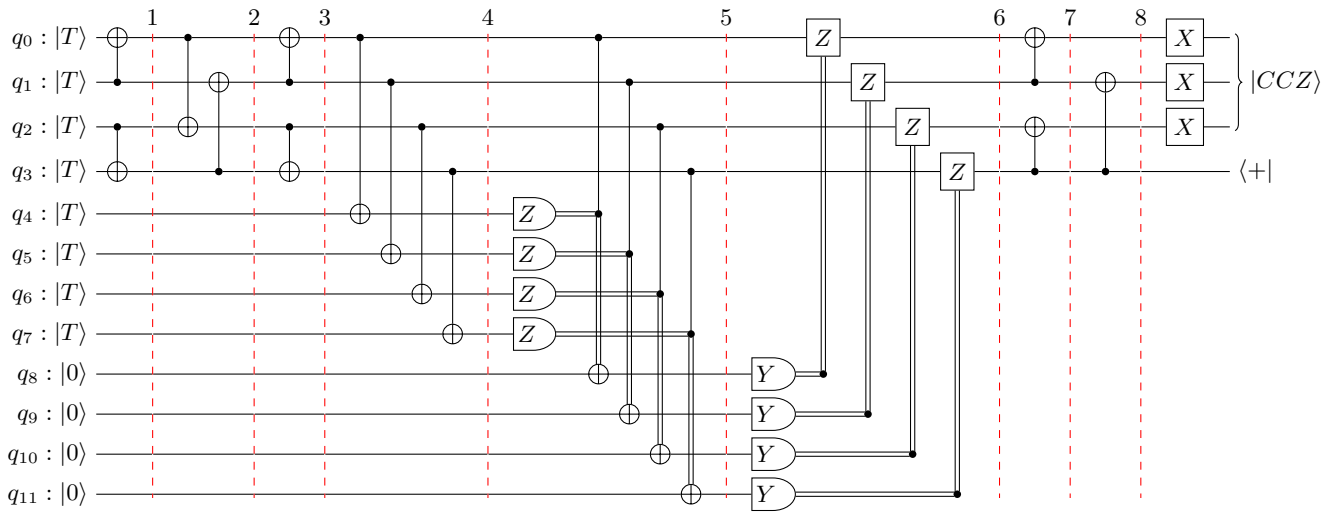

As a comparison, if we instead use the standard rotated surface code as proposed in \cite{PRXQuantum.4.020345}, the transversal $S$ gate is not available. In this case, we implement the $S$ gate via the $Y$-basis measurement shown in \cref{fig:tele_S}. With this modification, the corresponding $CCZ$ factory implementation is shown in \cref{fig:CCZ}. We note that an alternative approach to implementing the $S$ gate is the half-distance $S$ gate proposed in ref.~\cite{PhysRevX.7.021029}, which requires $d$ code cycles and has a higher error rate, but can be applied in parallel with magic state cultivation and other gates on $q_0$ to $q_7$. One can immediately see that the logical qubit overhead of the factory increases from 8 to 12. However, since each code patch is now single-layered (unfolded), all patches can still fit within the same stack, accommodating 12 qubits per loop. In fact, the increased qubit count reduces the cultivation runtime, from 22 code cycles to 15 code cycles. For $n = 12$, the resulting code cycle time, calculated as described in \cref{Sec:folded} and rounded up to allow for slack, is $5\mu s$. All CNOT gates remain transversal, and one round of stabiliser checks is performed at each time slice. The Y basis measurements that require $(0.5d+2)T_{cyc}^*(12)$ are now the bottleneck of this circuit. The runtime of the factory increased to $\sim 279\mu s$ for $d=25$. The overall spacetime volume, relative to that of \cref{fig:CCZ_folded}, increases by a factor of approximately 2.6. For completeness, we note that the conventional 8T-to-CCZ factory based on full lattice surgery requires $5d$ to $6d$ code cycles on 12 logical qubits \cite{gidney2025factor}, and is therefore substantially more costly.

The space and time overheads associated with logical Clifford operations and 8T-to-CCZ distillation for the rotated surface code without looped pipelines, the rotated surface code with looped pipelines, and the folded surface code with looped pipelines are summarized in \cref{tab:spacetime}. Runtime estimates assume silicon spin qubits where suitable, with a code-cycle time of \(3~\mu s\) for the standard rotated surface code. This value is chosen based on the analysis in \cref{sec:silicon_implem}, which yields \(T_{cyc}(n=2)=3.15~\mu s\). Space overhead here refers to the number of shuttling loops required, which corresponds to the actual physical real estate. A square surface code patch occupies unit area, while a triangular folded surface code patch has area \(0.5\). Note that we do not count the number of qubits per loop (i.e., the number of logical qubits per stack) toward the space cost. This is because introducing additional qubits into each loop via pipelining does not increase the number of physical devices required (for chip-based shuttling tracks) or the overall spatial footprint. Instead, the cost of adding qubits to existing loops manifests as an increase in the code-cycle time and is therefore accounted for in the time cost rather than the space cost.

\begin{table*}
    \centering
    \begin{tabular}{l|c|cccc}
         & & $H$         & $S$            & (intra-stack) CNOT & 8T-to-CCZ \\
         \hline
         & standard  & $3d\,T_{cyc}$ & $1.5d\,T_{cyc}$ & $2d\,T_{cyc}$ & $\sim5d\,T_{cyc}$ \\
         \parbox[t]{18ex}{Runtime} & pipelined rotated & $3d\,T_{cyc}$ & $1.5d\,T_{cyc}$    & $\sim1\mu s$ & $(d+27)T_{cyc}^*(12)+19\mu s$ \\
         & pipelined folded & $\sim T_{cyc}$ & $\sim T_{cyc}$ & $\sim1\mu s$ & $33T_{cyc}^*(16)+18\mu s$\\
         \hline
         & standard & $2$ & $2$ & $3$ & $12$ \\
         \parbox[t]{18ex}{Space}  & pipelined rotated & $2$ & $1$ & $1$ & $1$\\
         & pipelined folded & $0.5$ & $0.5$ & $0.5$ & $0.5$\\
         \hline
         \parbox[t]{18ex}{Spacetime saving of folded over...} & standard & $\sim12d$ & $\sim6d$ & $\sim36d$ & $\sim 1.667d$ \\[-8pt]
         &  pipelined rotated & \parbox{2.5cm}{$\sim12d$} & \parbox{2.5cm}{$\sim3d$} & \parbox{2.5cm}{$\sim2$} & \parbox{2.5cm}{$\sim 0.046d+1.426$}\\
    \end{tabular}
    \caption{Space and time overheads of logical gates and CCZ distillation for the standard rotated surface code (without looped pipelines), the rotated surface code with looped pipelines, and the folded surface code with looped pipelines. Runtime estimates assume silicon spin qubits where suitable, with a code cycle time of \(3~\mu s\) for the standard rotated surface code. Space overheads measure the required physical area, where a square surface code patch has unit area and a triangular folded surface code patch has area \(0.5\). The parameters of the CCZ factory are described in \cref{Sec:magic}. The final two rows report the improvement factors in spacetime volume achieved by the folded surface code relative to the other two approaches.}
    \label{tab:spacetime}
\end{table*}

\section{virtual stack architecture}
\label{Sec:architecture}

\begin{figure*}
    \centering
    \includegraphics[width=\linewidth]{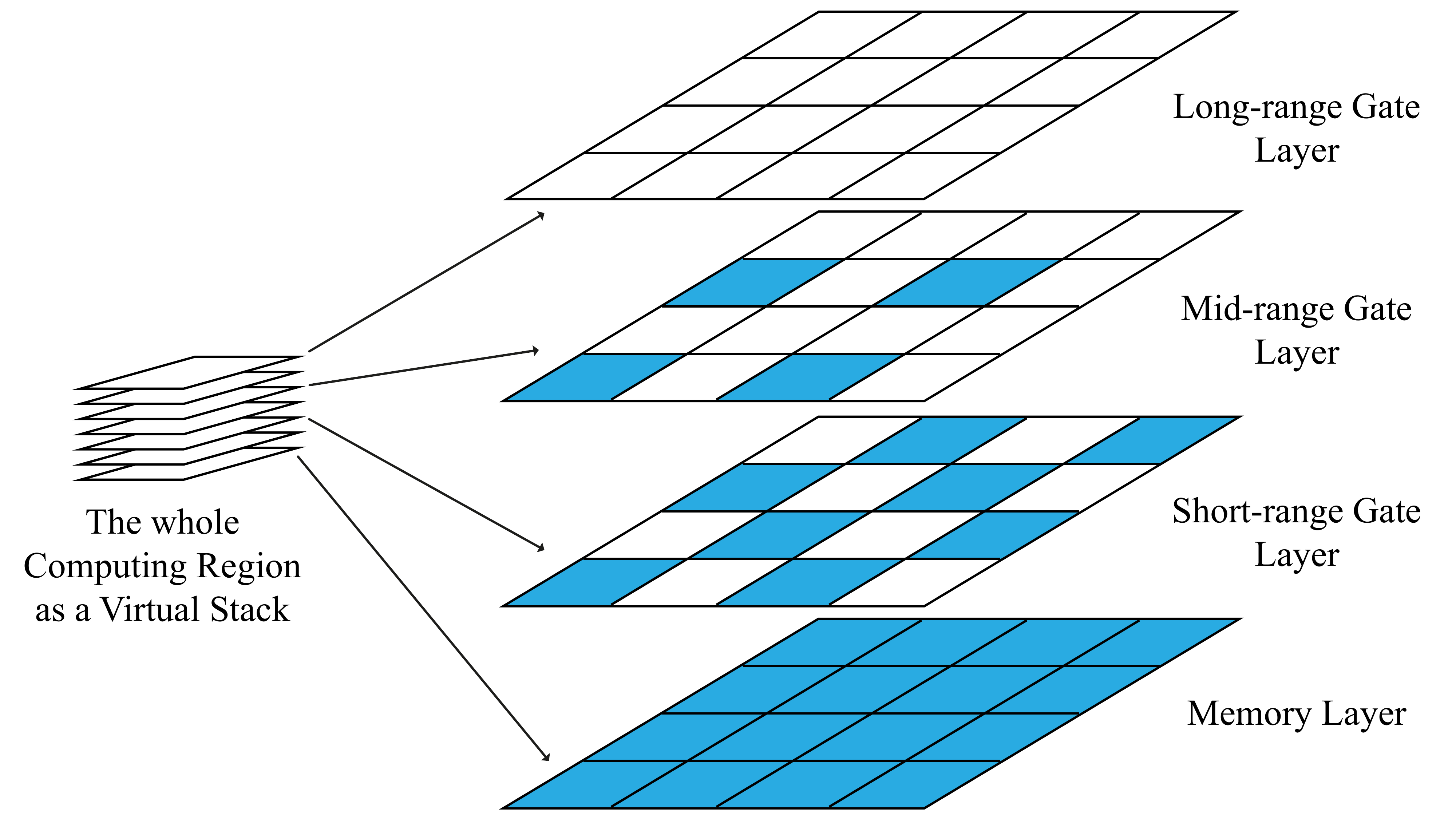}
    \caption{Schematic diagram illustrating how the computing region in a looped-pipeline architecture can be organized into layers with different connectivities. Four types of layers are shown, corresponding to memory, short-range gate, mid-range gate, and long-range gate operations. Blue and white code patches denote data and ancilla qubits, respectively.}
    \label{fig:layers}
\end{figure*}

The layout choice for our new triangular code patches can, of course, be the same as in \cref{fig:stacks_hallway}, except that now two stacks shaped as triangular prisms are joined together to form a rectangular stack. The ``hallways'' offer ample routing space, enabling efficient lattice surgery between the virtual stacks. This layout is simple and compatible with most algorithms, but a clear improvement would be to vary the routing space allocated to qubits performing different tasks. For example, a denser floor plan can be used for memory qubits, which are accessed less frequently. This idea of partitioning work regions according to connectivity requirements has also been discussed, for example, in ref.~\cite{gidney2025factor}.

\begin{figure}
    \centering
    \includegraphics[width=\linewidth]{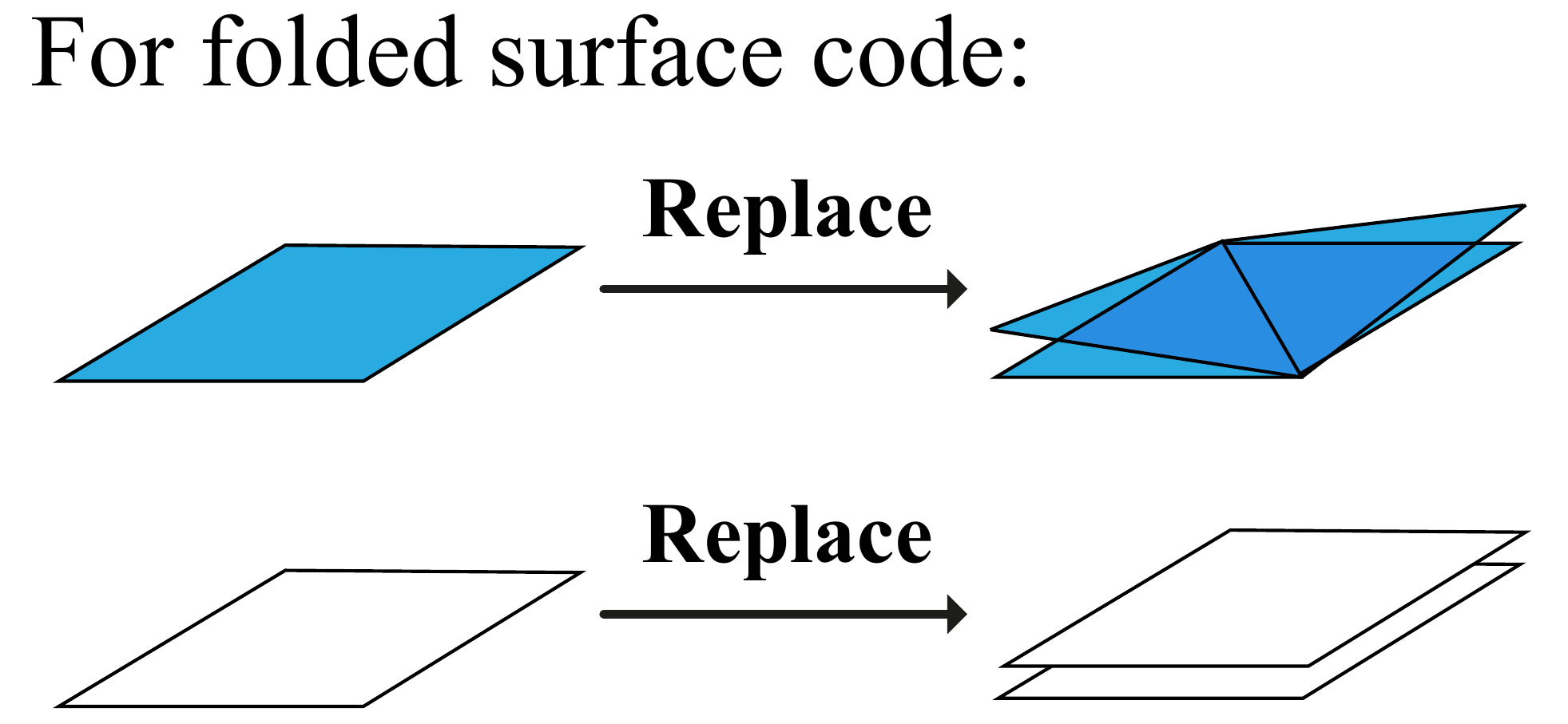}
    \caption{\black{Modification rules for \cref{fig:layers} when folded surface code patches are used: two triangular patches are combined to form a double-layer blue square, and each white square patch is also converted into a double-layer structure.}}
    \label{fig:Folded_rule}
\end{figure}

Within a looped pipeline architecture, this layout is realized by embedding routing space within each stack. This eliminates the need for separate stacks dedicated to data storage and routing ancilla, since each stack now contains both data qubits and internal routing capacity. Intuitively, this will give us a picture of layers as shown in \cref{fig:layers}. Different layers can be designed for different purposes. The memory layers store qubits with minimal routing space. The short-range gate layer may, for example, adopt a checkerboard layout to support localized interactions, which can be used as hot storage. The mid-range gate layer can employ the hallway layout to support parallelism, while the long-range gate layer is kept almost empty so that qubits can be swapped into it to facilitate long-range interactions. Note that although \cref{fig:layers} depicts code patches as squares, the same principles apply to more general codes, such as colour codes. In the case of the folded surface code, two triangular patches are paired to form a “double-layered” blue square in \cref{fig:layers}, while the white patches remain regular rotated surface code patches, albeit also double-layered. Consequently, each labelled layer in \cref{fig:layers} is effectively double-layered for the purpose of efficient lattice-surgery routing. \black{The modifications described above are summarized in \cref{fig:Folded_rule}.}

\begin{figure*}
    \centering
    \includegraphics[width=0.32\linewidth, trim=0cm 1.9cm 0cm 0cm, clip]{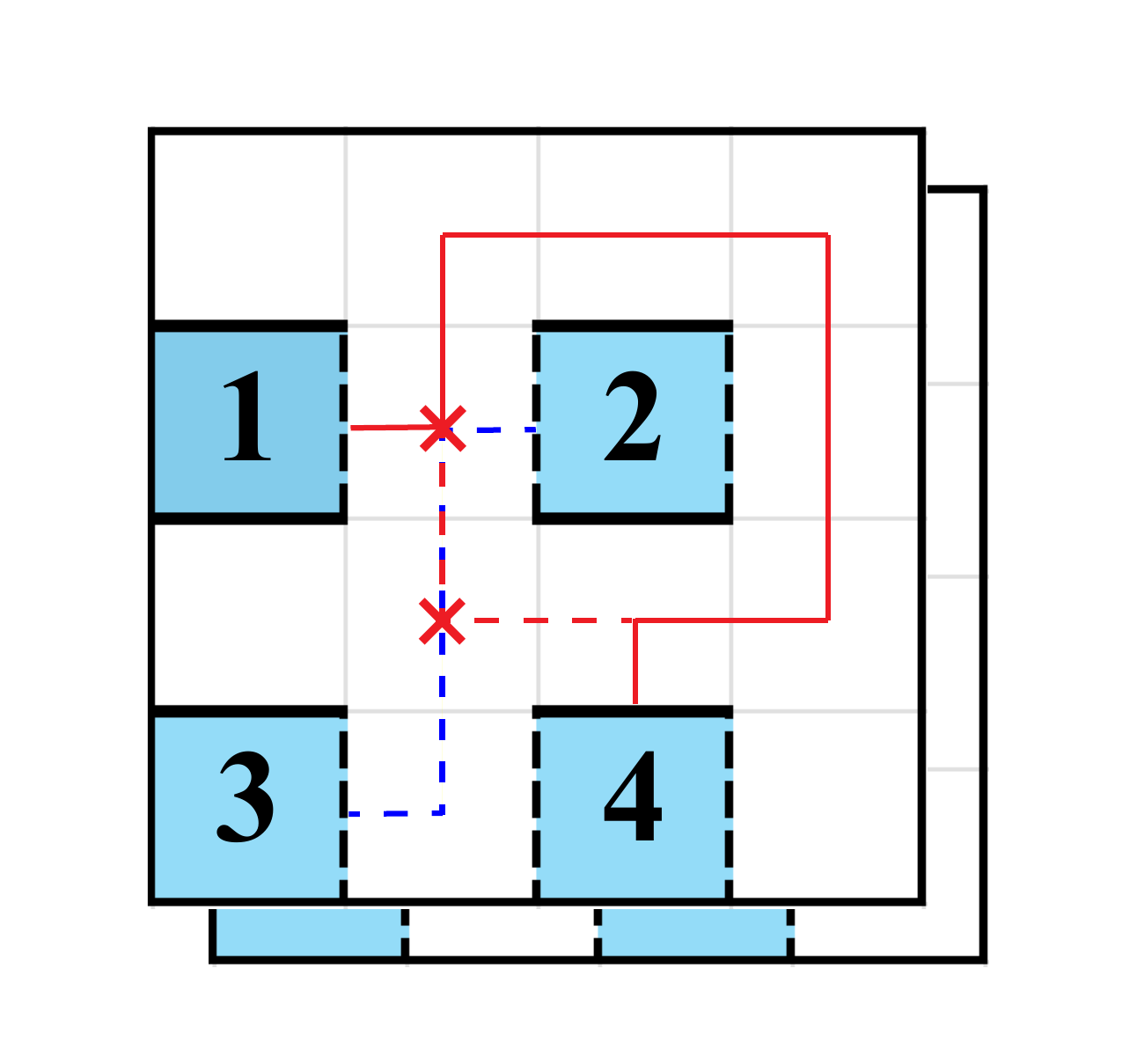}
    \label{fig:example_a}
    \subfloat[\justifying Eight logical qubit patches arranged into the hallway layout, resulting in 4 stacks of code patches, each contain 2 layers. In this layout, measuring $Z_1X_4$ and $Z_2Z_3$ in both layers simultaneously is impossible: no matter which path we choose to measure $Z_1X_4$ (solid or dashed red line), $Z_2$ is inaccessible to $Z_3$ in both layers.]{
    \includegraphics[width=0.9\linewidth, trim=0cm 19cm 0cm 5cm, clip]{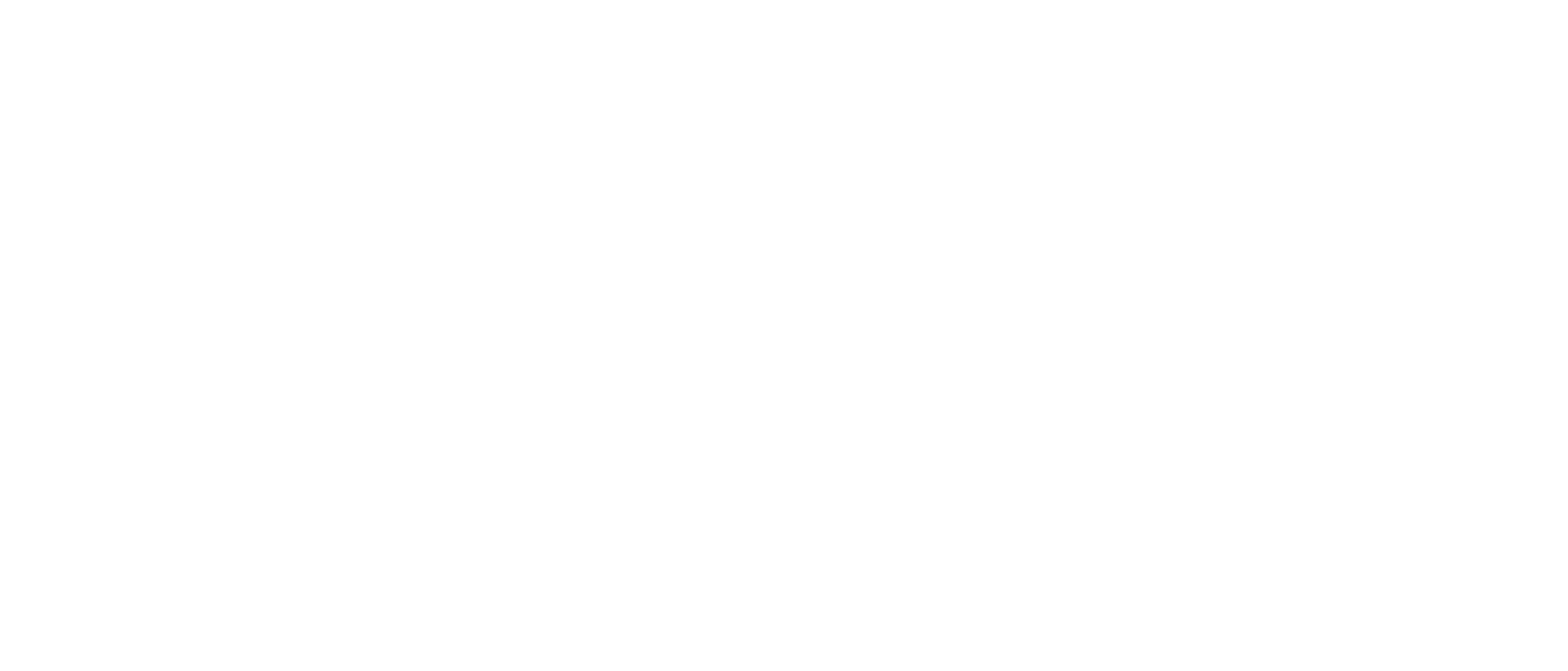}
    }\\
    \includegraphics[width=0.8\linewidth, trim=0cm 1cm 0cm 0cm, clip]{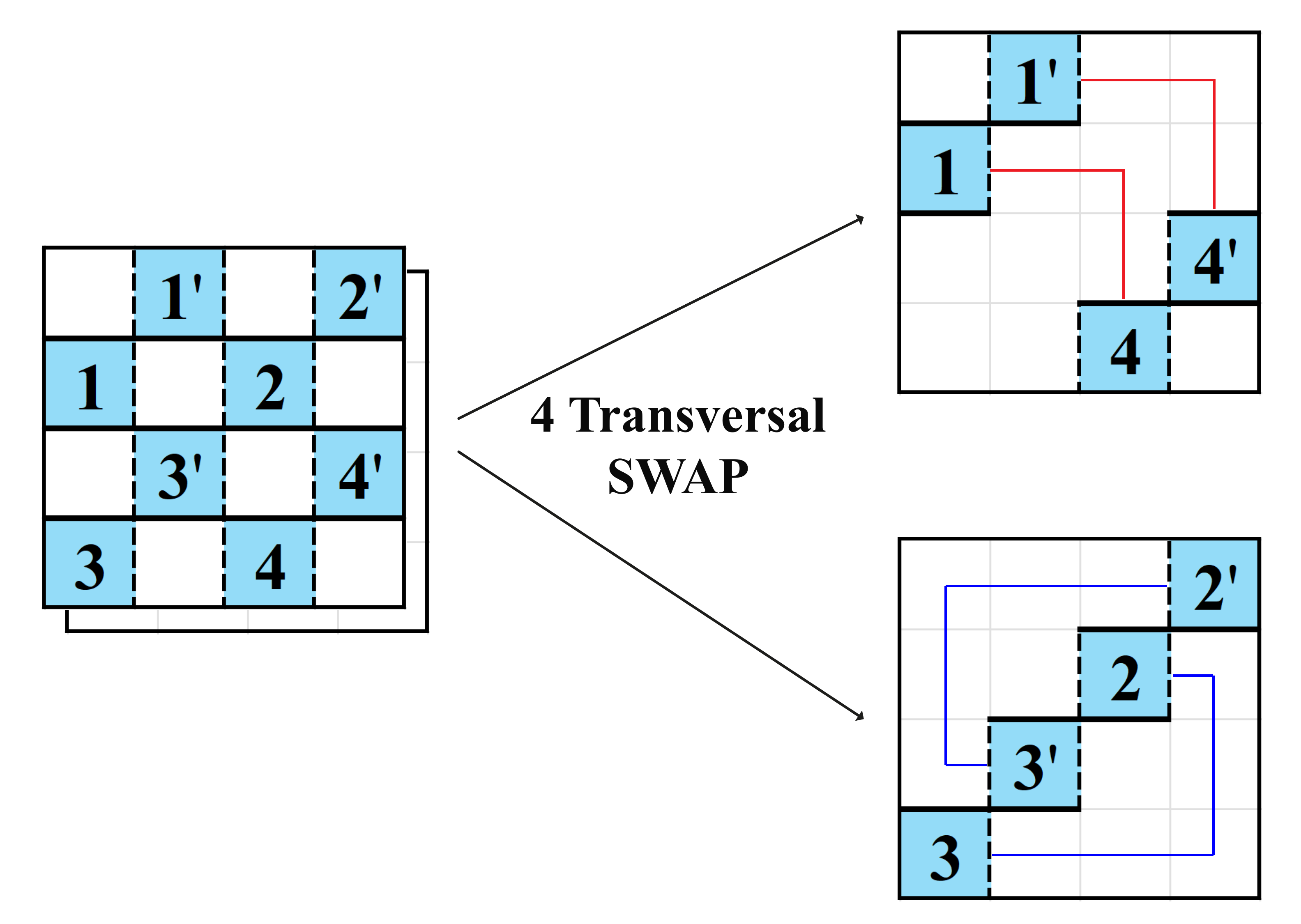}
    \label{fig:example_b}
    \subfloat[\justifying Eight logical qubit patches arranged into the checkerboard layout, plus a long-range gate layer (so that the average qubit density per layer is the same). With 4 transversal SWAP gates, all the $Z_1X_4$ and $Z_2Z_3$ can be measured simultaneously.]{
    \includegraphics[width=0.9\linewidth, trim=2cm 19cm 2cm 5cm, clip]{figures/blank.png}
    }
    \caption{An example illustrating that the new layout scheme shown in \cref{fig:layers} is unconditionally better than the hallway layout shown in \cref{fig:stacks_hallway}. Solid and dashed boundaries denote $X$ and $Z$ boundaries, respectively. Although rotated surface code patches are used for illustration, the same argument applies to folded surface code patches.}
    \label{fig:example}
\end{figure*}

It is evident that this layout is at least as efficient as the original hallway layout, since the latter can be regarded as a special case in which all layers correspond to the mid-range gate layer. Moreover, the new layout can be substantially more efficient by enabling parallelisation that are not possible in the hallway layout. To demonstrate this point, we consider the example shown in \cref{fig:example}, where rotated surface code patches are used for illustration, with the $X$ and $Z$ boundaries indicated by solid and dashed lines, respectively. Similar examples can be constructed for folded surface code patches. Imagine we have eight logical qubit patches arranged in the hallway layout, resulting in four stacks of code patches, each containing two layers. Now we want to measure $Z_1X_4$ and $Z_2Z_3$ in both layers (shown in \cref{fig:example_a}).However, irrespective of whether the solid or dashed red path is used to measure $Z_1X_4$, a simultaneous measurement of $Z_2Z_3$ is not possible. If we instead choose the checkerboard layout together with an empty long-range gate layer (the two cases have the same average qubit density per layer), then, using 4 transversal SWAP, we can measure all the $Z_1X_4$ and $Z_2Z_3$ simultaneously, as shown in \cref{fig:example_b}. And we have seen in \cref{Sec:folded} that transversal SWAP is almost free compared to the $O(d)$ runtime of lattice surgery. In fact, if we choose to SWAP code patches $1^\prime,2^\prime,3^\prime\text{ and }4^\prime$ to the empty layer, we have two layers in hallway layout. This shows that the layout scheme in \cref{fig:example_b} contains all the connectivity available in \cref{fig:example_a} and adds in more flexibility, with the aid of cheap transversal SWAP.

The memory layer can employ the yoked surface code \cite{gidney2025yoked} to achieve more space-efficient storage. The 2D yoked surface code is expected to be used as cold storage, where qubits are accessed infrequently; consequently, only limited routing space is allocated for the code patches. And it was envisaged in ref.~\cite{gidney2025yoked} that the surface code used for hot storage should be at most 1D yoked, but not 2D. However, taking advantage of the quasi-3D nature of the looped pipeline architecture, it may be possible to employ 2D yoked surface codes for hotter storage, thereby improving storage efficiency. With the help of cheap transversal SWAP and a long-range gate layer, the yoked code patches can be easily accessed for operations. The caveat is, however, the yoked code patches have smaller code distance than regular unyoked patches, which makes them more vulnerable to noise when exposed. Immediately after a yoked code patch is swapped into another layer, $O(d)$ code cycles might be required to first expand the code patch to regular size, whereas shrinking it when swap the patch back is almost free. Furthermore, the performance of operating on yoked surface code in large scale architecture is not yet well understood, and the detailed scheme of incorporating folded surface code with yoke checks remains to be established.

To fully exploit the advantage of the new layout in practice, the compiler must be aware of the degree of locality associated with the operations on each algorithmic qubit when performing qubit mapping. This will also help determine the portions of the four types of layers during compilation. In the ideal case, the compiler would determine the optimal occupancy of each layer and arrange the layout accordingly, rather than selecting from predefined layouts with fixed occupancies of $\{0\%, 25\%, 50\%, 100\%\}$, as outlined above. Moreover, if yoked surface code is employed, then the scheduler should be aware of the window between the yoke checks, during which the yoked code patches can be operated on.

\section{Conclusion}
\label{Sec:Conclusion}

In this work, we propose a scalable two-dimensional architecture based on looped pipelines for implementing folded surface codes. We present explicit protocols for performing all logical Clifford gates transversally within each virtual stack and show that this approach reduces the distillation overhead of the CCZ state by more than an order of magnitude compared with conventional methods, and by approximately a factor of 2.6 relative to prior looped pipelined surface code architectures. Furthermore, we introduce a new virtual stack layout scheme that exploits the layered structure of the architecture. This enables forms of gate parallelization that were previously inaccessible, providing a more efficient and flexible framework for compiling large-scale quantum computations.

We consider benchmarking the performance of looped pipeline architectures with folded surface codes to be a valuable direction for future work, including the development of a compiler optimized for this architecture. Moreover, as mentioned earlier, devising a detailed scheme for performing yoke checks on folded surface codes and evaluating their performance constitutes another promising avenue of investigation. Furthermore, although inter-stack operations still require lattice surgery, this work enables transversal logical Clifford gates within the same stack; it may therefore be possible to develop variants of the algorithmic fault tolerance concept introduced in ref.~\cite{zhou2025low}, potentially allowing further reductions in runtime. Finally, in our current design, the magic state factories employ the standard magic state cultivation protocol, which is tailored to square-grid connectivity. Several improved protocols have been proposed \cite{chen2025efficient,sahay2025fold,hirano2025efficient}, some of which assume long-range connectivity. Adopting these improved cultivation protocols, developing variants optimized for the looped pipeline architecture, or exploring alternative schemes for implementing magic gates using looped pipelines \cite{scrubyFaulttolerantQuantumComputation2025} are all compelling directions for future research.

\section*{Acknowledgement}

The authors thank Simon Benjamin, Zihan Chen, B\'alint Koczor and Adam Siegel for fruitful and inspiring discussions.
ZC acknowledges support from EPSRC projects Robust and Reliable Quantum Computing (RoaRQ, EP/W032635/1) and the EPSRC quantum technologies career acceleration fellowship (UKRI1226).
\appendix

\section*{AUTHOR CONTRIBUTIONS}

Z.S. and Z.C. conceived the study. Z.S. performed the detailed analysis and wrote the original draft of the manuscript. Z.C. supervised the project. Both authors reviewed and edited the manuscript.

\section*{COMPETING INTERESTS}

The authors declare no competing interests.

\section{Protocol Runtimes}
\label{appx:runtimes}

\subsection{Qubit rearrangement}
\label{appx:rearrangement}

\begin{figure*}
    \centering
    \includegraphics[trim={4cm 22cm 15.5cm 7cm},clip,width=0.9\linewidth]{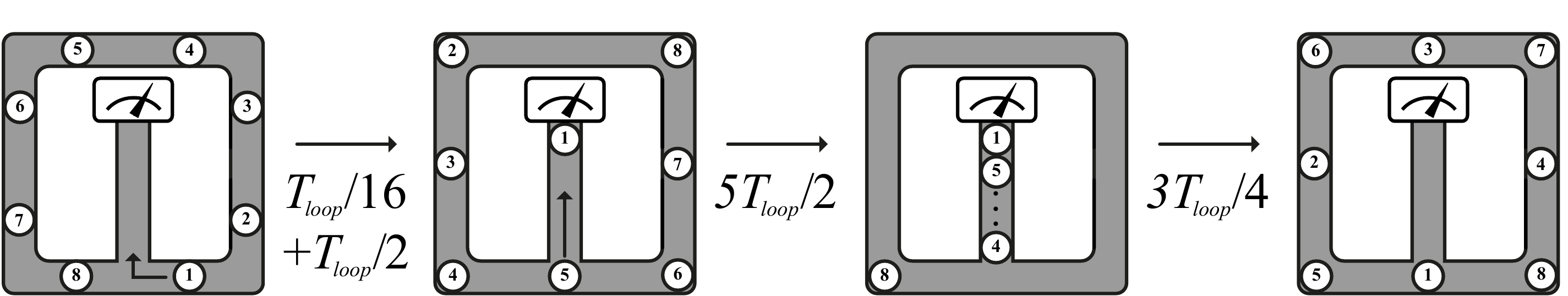}
    \caption{Example of qubit rearrangement in an eight-qubit loop. The qubits are reordered from the sequence \(1\,2\,3\,4\,5\,6\,7\,8\) (anticlockwise) to \(3\,7\,4\,8\,1\,5\,2\,6\). The initial configuration corresponds to a worst-case scenario. Circles indicate qubits and are not drawn to scale. The total duration of the process is $\frac{61}{16}T_{loop}$.}
    \label{fig:Rearrange}
\end{figure*}

Rearranging qubits within a single loop into an arbitrary ordering constitutes a more challenging intra-loop interaction task. We illustrate our rearrangement scheme with an example for \(n = 8\). Suppose the initial qubit configuration is as shown in the leftmost loop of \cref{fig:Rearrange}, corresponding to a worst-case scenario. Without loss of generality, the qubits are labelled \(1\)–\(8\) in the anticlockwise direction. For illustration, we aim to rearrange the qubits into the sequence \(3\,7\,4\,8\,1\,5\,2\,6\). Because the qubit ordering is cyclic, we may always shuttle first the qubit closest to the port entrance; in this example, the target permutation \(3\,7\,4\,8\,1\,5\,2\,6\) is therefore equivalent to \(1\,5\,2\,6\,3\,7\,4\,8\). In the worst case, shuttling this qubit into the port requires $\frac{1}{2n}T_{loop}$. The remaining qubits are then shuttled into the port in the desired order. Shuttling qubit 5 requires a time of $T_{loop}/2$, and then shuttling qubit 2 requires $(\frac{1}{2}-\frac{1}{n})T_{loop}=\frac{3}{8}T_{loop}$. The qubit pairs $(5,2)$, $(6,3)$ and $(7,4)$ each require a shuttling time of $(1-1/n)\ T_{loop}$, giving a total of $(\frac{n}{2}-1)(1-\frac{1}{n})\ T_{loop}=\frac{21}{8}T_{loop}$. Finally, the last qubit (qubit 8) need only be shuttled to within a distance of \(L_{loop}/n\) from the port entrance, as illustrated in the middle-right loop of \cref{fig:Rearrange}. This final step takes $(\frac{1}{2}-\frac{1}{n})T_{loop}=\frac{3}{8}T_{loop}$. The qubits in the port are subsequently shuttled back out to complete the rearrangement, which requires $(1-\frac{2}{n})T_{loop}$. Altogether, this process takes 
\begin{equation}
\begin{aligned} \label{eqn:rearrange}
&\Biggl[
\frac{1}{2n}
+ \left(\frac{n}{2}-1\right)\left(1-\frac{1}{n}\right)
+ \left(\frac{1}{2}
- \frac{1}{n}\right)
+ \left(1
- \frac{2}{n}\right)
\Biggr] T_{loop} \\
&= \left(\frac{n}{2}-\frac{3}{2n}\right) T_{loop} 
\end{aligned}
\end{equation}
which evaluates to $\frac{61}{16}T_{loop}$ for the \(n=8\) example, corresponding to $1.525\mu s$ for spin qubits. And for the scheme we developed, \cref{eqn:rearrange} represents the worst-case shuttling time required for rearrangement. We note that when \(n\) is odd, a similar derivation shows that the worst-case time scales slightly better, as $(\frac{n}{2}-\frac{2}{n})T_{loop}$.

A SWAP-based rearrangement strategy, where qubits are swapped pairwise to achieve the desired ordering, is slower in the worst case but can be faster when the target configuration closely resembles the current one. Consequently, both schemes are useful in practice for achieving optimal performance.

\subsection{Stabiliser checks for folded surface code} \label{appx:stabiliser}

\begin{figure*}
    \centering
    \includegraphics[width=0.95\linewidth, trim=9cm 4cm 5.5cm 3cm, clip]{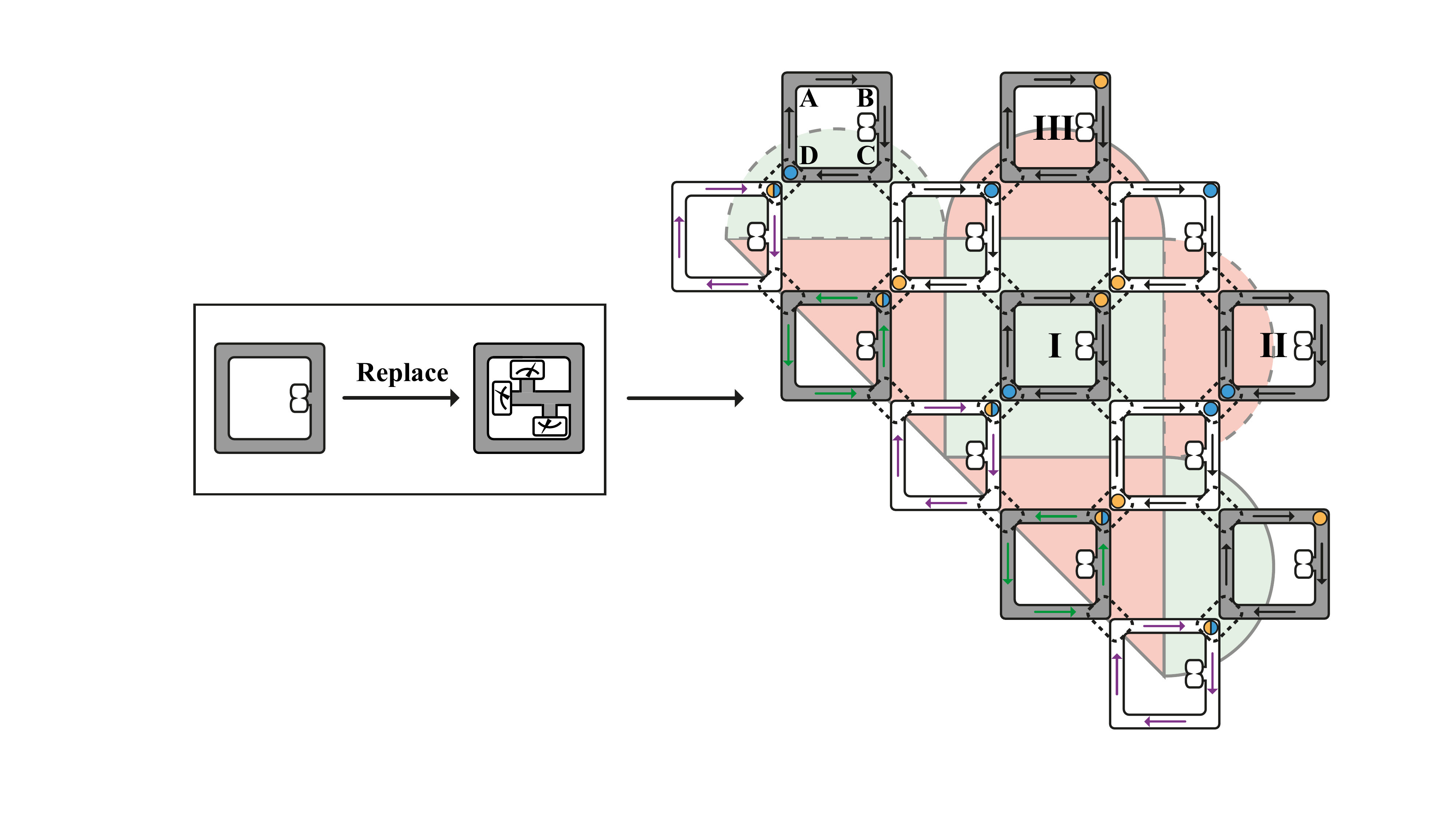}
    \caption{A single folded surface code patch with corners and loops labelled for explanatory purposes. All loops (both white and grey) should be understood as having the replacement rule shown on the left applied, such that each loop is equipped with three measurement devices.}
    \label{fig:explain}
\end{figure*}

In this appendix, we describe the stabilizer checks protocol in detail with reference to \cref{fig:explain}, where the four corners of the loops are marked A, B, C and D, and several loops are labelled with Roman numerals for later reference. Note that the order of the CNOT gates is described in \cref{Sec:folded}.

For simplicity and without loss of generality, we assume that at the beginning of the stabilizer cycle the yellow qubits have already been shuttled into the positions required for the first layer of CNOT gates in the check circuit. This configuration corresponds to shuttling qubits in off-diagonal loops by a distance of $L_{loop}/4$ (or $L_{loop}/2$ for qubits in diagonal loops) against the direction indicated by the arrows in the loops. Therefore, for example, the yellow qubit in loop I is located at corner A and the blue qubit is at corner C. The yellow qubits can then execute the first layer of CNOT gates, which takes a time $T_{2q}$. Subsequent shuttling is required to proceed: for instance, in loop I, the next operation is the first layer of CNOT at corner A for the blue qubit, requiring a shuttling time of $T_{loop}/2$ followed by a gate time $T_{2q}$. The second layer of CNOT gates for the yellow and blue qubits in loop I can be applied simultaneously at corners D and B, respectively. This step requires a time of $T_{loop}/4+T_{2q}$. Altogether, in loop I, the first two layers of CNOT takes a total time of $\frac{3}{4}T_{loop}+3T_{2q}$. However, the first two layers of CNOT gates in loop II constitute the time-limiting step. The blue qubit starts at corner C, with the first and second CNOT gates applied at corners A and D, respectively, resulting in a total time of $\frac{5}{4}T_{loop}+2T_{2q}$.

Once the first two layers of CNOT gates for all loops are completed (i.e., immediately after the second CNOT in loop II), the qubits occupy exactly the positions shown in \cref{fig:explain}. The remaining two layers of CNOT gates required for the stabilizer checks take an additional $\frac{5}{4}T_{loop}+2T_{2q}$. During this stage, loop III is the time-limiting loop, with the yellow qubit located at corner C after the final CNOT. Finally, the qubits are shuttled into the port between corners B and C for measurements. This step requires a time of $\frac{7}{8}T_{loop}$ for loop III, which is time-limiting loop. Assuming 3 measurement devices per loop and a measurement time $T_{meas}$, the measurement step is completed in a time $T_{meas}$. Accounting for the Hadamard gate time $T_{1q}$ at both the beginning and the end of the stabilizer checks, we obtain $T_{cyc}(n=2)=\frac{27}{8}T_{loop}+2T_{1q}+4T_{2q}+T_{meas}$. For spin qubits, this evaluates to $3.15 \mu s$.

\subsection{H, S and CNOT gate}

The protocols for the $H$ and $S$ gate are based on the stabiliser checks protocol. We again describe these protocols with reference to \cref{fig:explain}. 

Immediately after the first two layers of CNOT gates, the qubits occupy exactly the positions shown in \cref{fig:explain}. At this stage, we apply $S$ or $S^\dagger$ gates to the diagonal qubits and $CZ$ gates to the off-diagonal pairs. These two operations can be carried out in parallel, with the $CZ$ implementation using the intra-loop interaction protocol in \cref{Sec:rearrange} constituting the time-limiting step. We illustrate this process using loop I as an example. Shuttling the yellow qubit into the port requires a time of $T_{loop}/8$, followed by an additional $T_{loop}/2$ to shuttle the blue qubit into the port for interaction. The $CZ$ gate itself takes a time $T_{2q}$, and shuttling the qubits out requires an additional $T_{loop}/2$. At this point, the intra-loop interaction protocol is complete, with the yellow qubit positioned at the port entrance. However, the next layer of CNOT gates for the yellow qubit is located at corner B. We therefore shuttle the qubits for a further $T_{loop}/8$ in the direction opposite to the arrows. The stabilizer check protocol then proceeds as usual. Overall, these procedures add a runtime of $\frac{5}{4}T_{loop}+T_{2q}$ to the code cycle time $T_{cyc}$, yielding \cref{eqn:T_S}. The protocol for the $H$ gate, and the corresponding expression in \cref{eqn:T_H}, can be derived in an analogous manner.

\begin{figure}
    \centering
    \includegraphics[width=0.65\linewidth, trim=2.3cm 1.5cm 2.8cm 1.5cm, clip]{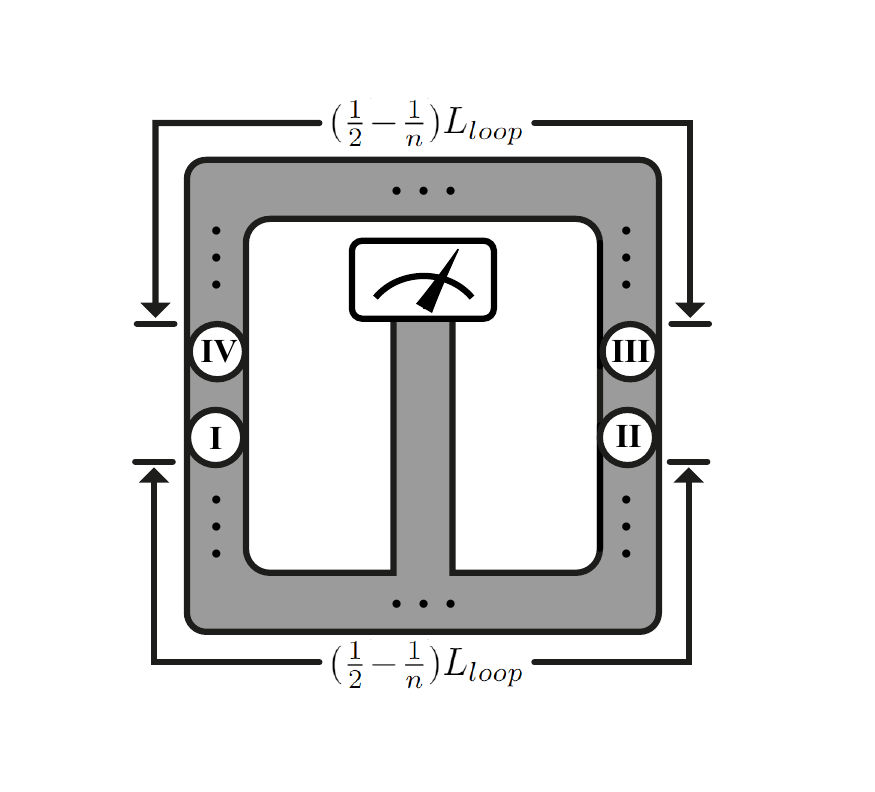}
    \caption{The worst-case scenario for the transversal CNOT protocol corresponds to applying $\mathrm{CNOT}_{I,II}$ and $\mathrm{CNOT}_{III,IV}$. The qubit pairs (I,II) and (III,IV) are each separated by a distance of $(\tfrac{1}{2}-\tfrac{1}{n})L_{loop}$.}
    \label{fig:cnot}
\end{figure}

The protocol for a transversal CNOT gate consists of applying the intra-loop interaction protocol twice. For a loop containing $n$ qubits (equivalently, $n/2$ code patches in a stack), the worst-case scenario corresponds to applying $\mathrm{CNOT}_{I,II}$ and $\mathrm{CNOT}_{III,IV}$ to the configuration shown in \cref{fig:cnot}. The qubit pairs (I,II) and (III,IV) are each separated by a distance of $(\tfrac{1}{2}-\tfrac{1}{n})L_{loop}$. In this case, shuttling qubit I into the port requires a time of $(\tfrac{1}{4}-\tfrac{1}{2n})T_{loop}$, while shuttling qubit II into the port requires $(\tfrac{1}{2}-\tfrac{1}{n})T_{loop}$. The subsequent gate operation and shuttling out together take $(\tfrac{1}{2}-\tfrac{1}{n})T_{loop} + T_{2q}$. Therefore, the total time required for $\mathrm{CNOT}_{I,II}$ is $(\tfrac{5}{4}-\tfrac{5}{2n})T_{loop} + T_{2q}$. At this point, qubit II is located at the port entrance, and qubit $III$ is only a distance $L_{loop}/n$ away. A similar analysis for $\mathrm{CNOT}_{III,IV}$ shows that it requires a time of $(1-\tfrac{1}{n})T_{loop} + T_{2q}$. The overall worst-case runtime is therefore given by \cref{eqn:T_CNOT}.

\subsection{8T-to-CCZ factories}
\label{appx:CCZ}

The runtime of the 8T-to-CCZ distillation circuit shown in \cref{fig:CCZ_folded}, implemented in a looped-pipeline architecture with folded surface code, is given by the following expression.
\begin{equation}
    T_{cul}+13 T_{CNOT}(16)+7 T_{cyc}+2 T_{meas}+4 T_S
\end{equation}

The runtime of cultivation is calculated as follows. We want to cultivate 8 $T$ states, each with output logical error rate $10^{-7}$, on 8 logical qubits with distance 25. From figure 1 of ref.~\cite{gidney2024magic}, we know the expected spacetime volume for cultivating a $T$ state with output logical error rate $10^{-7}$ is $\sim3 \times10^4$ qubit rounds. Therefore, the average cultivation time for 8 such $T$ states is
$$8\times\frac{3\times 10^4}{8\times2\times(25+1)^2}\sim22\text{ code cycles}$$

Note that, because the transversal CNOT scheme requires access to the port, CNOT operations that appear parallel in the abstract circuit, for example the four CNOTs between time slices 3 and 4, cannot be executed in parallel when there is only a single port in the loop. For the same reason, the four transversal $S$ gates also cannot be executed in parallel. Strictly speaking, the $4 T_S$ term in the runtime expression is an upper bound, as the four CNOT layers in the protocol can be streamlined to reduce some of the shuttling time. We also use the worst-case CNOT runtime given by \cref{eqn:T_CNOT} for all the CNOT gates. This approximation has negligible impact, since $T_{CNOT}$ is much smaller than the other runtimes.

For the four measurements on $q_4$ to $q_7$, since there are only 3 measurement devices per loop, so it requires $2 T_{meas}$ to complete the measurements. We neglect the time for decoding and other classical processing after the measurements.

The runtime of the 8T-to-CCZ distillation circuit shown in \cref{fig:CCZ}, implemented in a looped pipeline architecture with rotated surface code, is given by the following expression.
\begin{equation}
T_{cul}^\prime + 8 T_{cyc}+2 T_{meas}+17T_{CNOT}(12)+ 2\times(0.5d+2)T_{cyc}
\end{equation}

The cultivation time reduced by a factor of 1.5 since we now cultivate on 12 logical qubits instead of 8. The four Y basis measurements cannot be executed in parallel for there are only 3 measurement devices per loop.

\section{Looped pipeline architecture with inter-loop shuttling}
\label{appx:inter-loop}

In this appendix, we propose another potential improvement to the standard looped pipeline architecture. In what we discussed above, qubits within a loop always stay in the same loop. But now we connect the loops and allow inter-loop shuttling, which imposes higher demands on the hardware. While folded surface code enables cheap Clifford operations within a stack, we will see that inter-loop shuttling can improve inter-stack operations. We denote the time taken to shuttle a qubit from a data loop to the nearest data loop (connected by an ancilla loop, see e.g., \cref{fig:fold1}) as $T_{int}$, and one can expect $T_{int}\sim T_{loop}/2$. To isolate and analyse the effect of inter-loop shuttling, we assume the use of the standard surface code in the following subsections, unless otherwise specified.

\subsection{Hadamard gate}

In the standard looped pipeline architecture, patch rotation is implemented directly using lattice surgery. However, if inter-loop shuttling is permitted, patch rotation can instead be achieved through physical rotation of the patch by shuttling the physical qubits so that the patch is rotated by $90^\circ$. This operation may be subject to scratch errors induced by static defects \cite{494s-jd8h}. In most cases, such errors are benign or can be corrected, but in certain situations they may lead to logical errors. If the location of the defect is known in these cases, performing the patch rotation in the opposite direction can avoid the error.

The time taken for this process is dictated by the path travelled by the physical qubits at the patch boundaries, they need to be shuttled through $d-1$ data loops. Therefore the time taken for this operation is $(d-1)T_{int}$.

An alternative approach combines inter-loop shuttling with lattice surgery. In lattice-surgery-based patch rotation, the final $d$ code cycles are used to move the rotated patch from the ancilla back to its original position; this step can instead be implemented via shuttling. The effect of scratch errors in the $Z$ basis can be mitigated by aligning the shuttling direction with the logical $X$ operator. This patch-rotation method, and hence the implementation of a Hadamard gate, requires one ancilla patch and $2d$ code cycles.

\subsection{SWAP, CNOT and CZ gate}

The SWAP gate between different stacks also benefits from inter-loop shuttling. In the standard looped pipeline architecture, inter-stack SWAP gate is implemented by patch movement via lattice surgery. With inter-loop shuttling, inter-stack SWAP can be implemented by directly swapping the corresponding physical qubits. By putting the two patches to be swapped in different layers of the stack, the shuttling of the patches can be parallelised. Similarly, this operation can be made to be resilient to the scratch noise in $Z$ basis. This SWAP requires no ancilla and time $d\,T_{int}$.

The inter-stack CNOT can be made cheaper now if implemented by inter-stack SWAP and transversal CNOT within the same stack. Therefore, the inter-stack CNOT can now be done in $\sim2d T_{int}$ and require no ancilla. With the more efficient Hadamard gate introduced in the previous section, $CZ_{ij} = H_j\,CNOT_{ij}\,H_j$ within/between stacks is also less costly now.

\subsection{S gate}

\begin{figure}[h]
\begin{tikzpicture}
\begin{yquant*}
qubit {$\ket{\psi}$} a;
qubit {$\ket{i}$} b;

cnot b|a;
h b;
cnot b|a;

output {$S\ket{\psi}$} a;
output {$Z\ket{i}$} b;

\end{yquant*}
\end{tikzpicture}
\caption{A teleportation-based scheme for implementing the $S$ gate. The unwanted $Z$ flip on the $\ket{i}$ state can be tracked in Pauli frame.}
\label{fig:S_magic}
\end{figure}
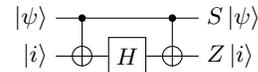

The final missing Clifford generator is $S$ gate. The choice for performing $S$ gate in the standard looped pipeline architecture is via Y basis measurement, which takes $1.5dT_{cyc}$. However, since the Hadamard can now be executed in $\sim(d-1)T_{int}$, with $T_{int}$ being an order of magnitude smaller than $T_{cyc}$, we can implement the $S$ via $\ket{i}$ teleportation using the circuit shown in \cref{fig:S_magic}~\cite{gidney2017slightly}. Assuming a $\ket{i}$ state is available in every stack, the cost for $S$ gate is dominated by the cost of $H$ as the CNOT gates are transversal. Moreover, if inter-loop shuttling is available, not only the Hadamard gates become cheaper but several stacks can share one $\ket{i}$ state, as SWAP is also cheap now.

\bibliography{bib}

\end{document}